\documentclass[10pt,twocolumn]{article}
\usepackage[T1]{fontenc}
\usepackage[utf8x]{inputenc} \PrerenderUnicode{àé}
\usepackage[francais]{babel}
\usepackage{JCAA}

\usepackage{ucs}
\usepackage{amsmath}
\usepackage{amsfonts}
\usepackage{amssymb}
\usepackage[numbers]{natbib}
\usepackage{url}
\usepackage{subfigure}
\usepackage{stfloats} 
\usepackage{relsize}
\usepackage{placeins}

\usepackage{graphicx}
\DeclareGraphicsExtensions{.pdf}
\graphicspath{{./images/}}

\title{Objets sonores: Une représentation bio-inspirée, hiérarchique, parcimonieuse à très grandes dimensions utilisable en reconnaissance}
\author{%
    Simon\ Brodeur et Jean Rouat\\
    Simon.Brodeur@usherbrooke.ca, Jean.Rouat@usherbrooke.ca\\
    Groupe de recherche en Neuroscience Computationelle et Traitement Intelligent des Signaux (NECOTIS) \\%
    Département g\'enie \'electrique et g\'enie informatique,
    Université de Sherbrooke,
    Sherbrooke QC Canada J1K 2R1\\
}

\begin{document}
\twocolumn[

\clearpage\maketitle
\thispagestyle{empty}

\begin{abstract}{ABSTRACT}
The emphasis is put on the hierarchical structure, independence and sparseness aspects of auditory signal representations in high-dimensional spaces, so as to define the components of auditory objects. The concept of an auditory object and its neural representation is introduced. An illustrative application then follows, consisting in the analysis of various auditory signals: speech, music and natural outdoor environments. A new automatic speech recognition (ASR) system is then proposed and compared to a conventional statistical system. The proposed system clearly shows that an object-based analysis introduces a great flexibility and robustness for the task of speech recognition. The integration of knowledge from neuroscience and acoustic signal processing brings new ways of thinking to the field of classification of acoustic signals.
\end{abstract}

\vspace{-1.5em}
\begin{abstract}{SOMMAIRE}
L'accent est placé dans cet article sur la structure hiérarchique, l'aspect parcimonieux de la représentation de l'information sonore, la très grande dimension des caractéristiques ainsi que sur l'indépendance des caractéristiques permettant de définir les composantes des objets sonores. Les notions d'objet sonore et de représentation neuronale sont d'abord introduites, puis illustrées avec une application en analyse de signaux sonores variés: parole, musique et environnements naturels extérieurs. Finalement, un nouveau système de reconnaissance automatique de parole est proposé. Celui-ci est comparé à un système statistique conventionnel. Il montre très clairement que l'analyse par objets sonores introduit une grande polyvalence et robustesse en reconnaissance de parole. Cette intégration des connaissances en neurosciences et traitement des signaux acoustiques ouvre de nouvelles perspectives dans le domaine de la reconnaissance de signaux acoustiques.
\end{abstract}


%
]

\maketitle

%

\section{Introduction}
L'organisation du système auditif reflète les structures des signaux sonores.
L'accent est placé dans cet article sur une proposition de représentation par objets sonores qui vise à intégrer certaines connaissances de la physiologie et de la perception dans la conception des objets sonores.
\subsection{Qu'est-ce qu'un objet sonore?}
Quels sont les structures et objets sonores à percevoir? Comment ces structures ou objets sont-ils identifiés par le système auditif? Comment mettre en oeuvre des systèmes de classification ou de reconnaissance capables d'extraire ou de reconnaître les objets sonores? Toutes ces questions sont encore ouvertes et non résolues, cependant, plusieurs observations physiologiques et psycho-acoustiques permettent de circonscrire les réponses potentielles par l'élaboration de modèles. Une fois les réponses circonscrites, il est intéressant de confronter ces modèles avec les outils des traitements de signaux, de la théorie de l'information et de l'intelligence artificielle, afin d'affiner notre compréhension de la perception d'objets sonores et de leur représentation. Par cet article, nous répondons partiellement en proposant une représentation simple des objets sonores et nous indiquons comment il est possible d'utiliser les connaissances des neurosciences pour proposer un modèle de traitement des sons afin de réaliser une analyse des signaux qui conduit à une reconnaissance robuste et immunisée contre les perturbations extérieures. Nous posons l'hypothèse que l'architecture du système auditif est aussi liée à la structure des objets sonores, c'est-à-dire que l'évolution a fait en sorte que les structures respectives des objets sonores et l'architecture du système auditif sont intimement liées. Nous pouvons alors transférer une partie des connaissances des neurosciences vers le traitement du signal pour élaborer de meilleurs systèmes.

\subsection{Tentative de définition d'un objet sonore}
Un objet (ou entité) est une structure indépendante (autonome) dont les composantes sont liées ensemble comme faisant partie d'un même objet. Il est donc possible de manipuler les objets indépendamment les uns des autres. Cependant, une modification sur une composante aura un impact sur tous les objets élaborés à partir de cette composante. Nous considérons les objets sonores qui ne peuvent être observés directement dans le signal acoustique en raison de son caractère spatio-temporel multi-échelle. À priori, une segmentation directe du signal continu introduira des erreurs d'estimation des objets sonores. En effet, les objets sonores ne sont séparables qu'une fois leurs caractéristiques perceptives obtenues. La séparation ne peut se faire que dans l'espace de la représentation auditive.

Notre environnement sonore est constitué d'éléments (composantes) que nous pouvons considérer comme indépendants, mais qui peuvent être assemblés (et donc liés) pour créer un objet, une partie d'objet ou une entité de niveau supérieur. Ces objets (ou entités) peuvent alors être manipulés de façon unitaire et indépendamment  les unes des autres. Un environnement sonore est donc le résultat de la combinaison de composantes élémentaires (construites indépendamment les unes des autres, avec éventuellement des contraintes externes de conception) qui sont ensuite assemblées et liées pour créer des objets plus complexes. Au sein de ces objets, les composantes élémentaires sont liées tandis que les objets qui sont cette fois-ci plus complexes peuvent être indépendants.

 Nous postulons aussi que ces parties élémentaires contribuent significativement à la perception sous forme d'évènements acoustiques de type ONSET, OFFSET, clics, ou d'unités plus stables de type modulations en amplitude (AM) ou en fréquence (FM), etc. L'état des connaissances actuelles ne permet pas de connaître à priori de façon exacte ces parties élémentaires, donc nous tentons de les approcher. Un même évènement sonore, c'est-à-dire dont les caractéristiques spatio-temporelles sont données, peut évoquer une perception différente suivant le bagage culturel et linguistique d'une population de personnes (e.g. occident, orient, langues tonales) ainsi que le contexte acoustique. Il est donc logique pour le concepteur de systèmes de reconnaissance de signaux acoustiques d'élaborer une stratégie permettant d'adapter la recherche de ces unités acoustiques élémentaires en fonction du contexte de l'utilisation du système. Une fois ces unités élémentaires trouvées il sera possible de les utiliser pour bâtir les représentations plus complexes des objets que nous souhaitons rechercher dans le signal.
 
 Nous intégrons dans le système proposé une recherche automatisée des unités élémentaires, puis de leur organisation afin de générer des objets sonores composés d'une organisation hiérarchisée des unités élémentaires. Nous donnons ci-dessous la démarche qui nous permet de procéder ainsi.

\subsection{Comment trouver les unités élémentaires acoustiques?}
Comment trouver ces unités ne connaissant pas à priori leurs formes ni leurs caractéristiques exactes, mais en disposant toutefois d'une quantité suffisante de signal acoustique? Dans ce contexte, il n'est pas possible d'utiliser une technique d'apprentissage utilisant des données à priori étiquetées (puisque les caractéristiques exactes de ce que nous cherchons ne sont pas connues). De plus, le nombre à priori de ces unités élémentaires n'est pas connu. Nous indiquons ci-dessous comment la connaissance du système auditif nous a orientés vers la solution choisie et qui permet de contourner ces difficultés. Pour cela nous exploitons des caractéristiques connues du cerveau:
\begin{enumerate}
\item L'accroissement de l'indépendance des activités neuronales lorsque l'on se déplace le long du chemin auditif, des noyaux périphériques vers le cortex auditif~\cite{chechikNeuron2006};

\item L'organisation hiérarchique du système auditif;

\item  La parcimonie neuronale et la très grande dimensionnalité des caractéristiques~\cite{molotchnikoffFrontiers2011}.
\end{enumerate}

\subsubsection{Champs récepteurs et unités élémentaires}
D'une certaine façon, le champ récepteur d'un neurone peut représenter une unité élémentaire acoustique (en réalité plusieurs neurones travaillant ensemble peuvent être interprétés en terme de champs récepteur). Plusieurs auteurs considèrent d'ailleurs qu'il y a analogie entre le filtre adapté (\textit{matching filter}) et l'opération de filtrage et de reconnaissance réalisée par un neurone. On peut citer par exemple le travail de Daniel L.~Alkon~\cite{alkon1990} qui propose en 1990 un modèle de la mémoire basé sur les caractéristiques physiologiques de neurones de l'escargot. Ce modèle considère que le champ récepteur d'un neurone s'adapte de façon à apprendre les relations de corrélations et d'anti-corrélations entre ses entrées et sa sortie. Le neurone est alors en mesure de ne répondre qu'à des configurations spécifiques de potentiels d'actions sur ses entrées. Ce thème a d'ailleurs orienté bon nombre de travaux de recherches actuels du domaine des neurosciences computationnelles. On peut citer au niveau du système auditif les travaux de l'équipe de Jos Eggermont~\cite{Chen2007} et de Shamma~\cite{depireux2001,zotkin2005}, qui considèrent le lien de  corrélation entre stimuli et champs récepteurs spatio-temporels de neurones. 
Bref, le champ récepteur d'un neurone auditif peut-être considéré comme étant équivalent en terme de traitement des signaux à une fonction dite de \emph{base}~\cite{depireux2001}. Ceci est aussi appuyé par les travaux de Lewicki~\cite{lewicki2002,Smith2006} au début des années 2000, qui constate que la représentation optimale des sons naturels passe par une projection de ceux-ci sur des bases dont la forme se rapproche des réponses des filtres cochléaires de l'audition~\cite{irino1997}. Il est d'ailleurs intéressant de constater que les psycho-acousticiens avaient observé ce type de réponse dès le milieu des années 1970 (e.g. \cite{Patterson1976}) et que ces mêmes réponses ont ensuite pu être obtenues uniquement à partir de critères d'optimisation de la représentation des sons~\cite{Smith2006}. 

Par ailleurs, il est connu depuis la fin des années 1980 que l'organisation des champs récepteurs  est sur-complète et que les neurones sont organisés en couches de caractéristiques spécifiques dans le noyau cochléaire~\cite{frisina1985}, pour le codage de l'information en modulation d'amplitude dans le colliculus inférieur\cite{Schreiner1988,Winer2005} et dans le cortex~\cite{schreiner1986}.
D'une certaine façon, les champs récepteurs peuvent être interprétés comme étant des bases parcimonieuses organisées de façon hiérarchique pour une représentation adaptée aux signaux auditifs. Il est possible d'exploiter cette architecture pour proposer de nouvelles façons de représenter les signaux sonores~\cite{Klein2003}. Assez tôt, il a été proposé de trouver automatiquement des bases surcomplètes en tenant compte du fait qu'il devrait y avoir une indépendance statistique entre les bases~\cite{Lewicki2000a}.

En résumé, on peut poser l'hypothèse que les bases sont indépendantes et perceptivement significatives, et que l'analogie avec les champs récepteurs de micro-circuits neuronaux du système auditif est possible. En utilisant ces critères, il sera possible de représenter les éléments ou parties d'objets sonores et permettre la conception d'un algorithme de recherche automatique de ces bases. En effet, les techniques actuelles en traitement des images disposent de plusieurs outils utilisables pour la recherche et le traitement d'objets sonores. Par exemple l'analyse en composante indépendante (ICA)~\cite{hyvarinen2001} permet de trouver des bases indépendantes et la factorisation en matrice non négative (NMF)~\cite{Lee1999,Donoho2004} des parties d'images qui sont interprétables visuellement.

\subsubsection{Hiérarchie du système auditif}
Un autre aspect très important à prendre en compte dans le traitement des sons est l'organisation très hiérarchisée du système auditif~\cite{Hickok2007}.
Il comprend de l'ordre de 7 à 8 noyaux nerveux traversés par l'information sonore avant de parvenir au cortex auditif. De plus, certains noyaux nerveux (e.g. le noyau cochléaire et le colliculus inférieur) sont aussi hiérarchisés. Cette organisation permet de prendre en compte le contexte acoustique~\cite{Lewicki1996}, soit la distribution spatio-temporelle des objets sonores. Elle permet aussi d'analyser de façon multi-échelle le signal acoustique, d’accroître la robustesse aux interférences et d'introduire une invariance dans la ``forme'' de la représentation multi-échelle et spatio-temporelle des objets sonores. De plus, une représentation mentale d'un objet sonore pourrait être élaborée par appariement de différentes couches (ou groupes) de neurones. Cet appariement pourrait se faire par synchronie des décharges des neurones (\emph{binding}). Ceci est d'ailleurs analogue à ce qui est observé dans le système visuel. L'hypothèse y est faite que la représentation mentale d'un objet puisse être le résultat de synchronie des décharges de sous groupes de neurones, chacun des sous-groupes représentant des caractéristiques ou parties différentes des objets~\cite{molotchnikoffFrontiers2011,visualCortexIntech2012}.

Le présent article présente une solution possible qui prend en compte la représentation objet des signaux telle que nous l'avons évoquée.
 Nous regardons maintenant comment la parcimonie et la très grande dimensionnalité des représentations peuvent être intégrées au traitement des signaux sonores.
%
%
%
%
\subsection{La parcimonie et la très grande dimension}
Peu de neurones sont actifs en même temps et leur réponse est parcimonieuse. Ceci est une conséquence indirecte du nombre considérable de neurones~\cite{molotchnikoffFrontiers2011}, qui conduit à une parcimonie spatiale. De plus, le premier neurone à répondre rapidement suite à la présentation d'un stimulus est celui qui encode et caractérise le mieux le stimulus. Il n'est alors pas nécessaire d'attendre la réponse des autres neurones~\cite{guyonneau2005NECO} et la parcimonie est alors aussi temporelle~\cite{guyonneau2004JouPhysiol}. En raison de cette parcimonie spatiale et temporelle, le codage de l'information repose sur des évènements discrets (décharges des neurones) distribués dans un espace à très grande dimensionnalité (i.e. où chaque neurone correspond à une dimension). Cet aspect est aussi pris en considération dans le système de reconnaissance qui est proposé à l'aide d'un codage binaire des caractéristiques.

\section{Bref résumé de l'état de l'art en analyse/reconnaissance de signaux sonores}
La vaste majorité des systèmes contemporains de reconnaissance/classification des signaux utilise le même type de caractéristiques pour représenter les signaux. Il s'agit des coefficients cepstraux sur l'échelle de Mel (MFCC) ~\cite{davis1980}. Leur utilisation est motivée par le fait que l'échelle des fréquences y est transformée en échelle Mel afin de reproduire la distribution des bandes critiques de l'oreille. Par ailleurs, la transformation en cosinus appliquée sur le logarithme du spectre d'amplitude, pour obtenir ces coefficients, reproduit dans une moindre mesure les patrons de connectivité par inhibition latérale tels qu'ils pourraient exister au niveau du noyau cochléaire. L'opérateur de logarithme permet ici la séparation entre la source glottale et le conduit vocal.
Mais le plus grand atout des MFCC est sans aucun doute le fait qu'en raison de la transformation en cosinus, ceux-ci sont relativement décorrélés, ce qui permet de meilleures performances lorsque le système de reconnaissance est basé sur les distributions statistiques des MFCC (e.g. les systèmes à base de chaînes de Markov~\cite{rabiner1989}).

De nouvelles architectures ont émergé depuis les dernières années, avec des résultats prometteurs. Ces nouveaux systèmes sont organisés de façon hiérarchique et mettent l'accent sur l'obtention de caractéristiques plus proches des propriétés perceptives du signal acoustique. On peut donner comme exemple~\cite{valenteINTERSPEECH2009,Mohamed2009,Mohamed2010,Seide2011}. Encore plus récemment, il a été démontré que la recherche de caractéristiques par l'utilisation d'architectures de réseaux de neurones à recherche profonde (Deep Belief Neural Networks,~\cite{hintonIEEESPM2012}) offre un potentiel supérieur aux systèmes de reconnaissance communément utilisés pour la parole spontanée. Le présent travail s'inscrit plutôt dans ce contexte.

%

\section{Approche proposée}
Nous proposons une nouvelle architecture qui prend en compte les points développés aux sections précédentes afin d'extraire une représentation des objets sonores qui respecte les propriétés importantes connues du système auditif. Nous le faisons en intégrant à la démarche des outils développés initialement pour le traitement des signaux et d'images afin de trouver, dans un premier temps, les bases surcomplètes qui seront équivalentes à des champs récepteurs placés à des niveaux hiérarchiques différents. Nous présentons une première expérience qui montre que la technique permet d'extraire effectivement des bases parcimonieuses caractéristiques des signaux utilisés durant l'apprentissage. Ensuite nous effectuons une reconnaissance de parole à l'aide de représentations parcimonieuses et de grandes dimensions. Dans notre cas, la grande dimension des données nous permet de travailler avec un codage binaire, ce qui permet d'accroître grandement la rapidité du traitement. Nous introduisons alors une reconnaissance à base de distributions de Bernoulli, adaptées à ce type de caractéristiques. Tous ces éléments permettent d'élaborer un système de reconnaissance de parole innovant.

Dans cette approche, les composantes élémentaires des objets sont les bases qui ont été trouvées pour le premier niveau hiérarchique. Les parties d'objets sonores (i.e. composantes complexes) sont représentées par les niveaux hiérarchiques supérieurs. Les objets ou parties d'objets sont représentés par des vecteurs dont les composantes sont binaires. Chaque coordonnée d'un vecteur objet (ou vecteur ``partie d'objet'') correspond à une composante particulière.
Ces composantes sont soit élémentaires (premier niveau de la hiérarchie), soit complexes (dernier niveau). Une composante (ou partie d'objet) est considérée comme appartenant à un objet sonore si la coordonnée du vecteur qui lui est associée est différente de zéro.



\begin{figure}[htb!]
\centering
\includegraphics[width=1.0\linewidth]{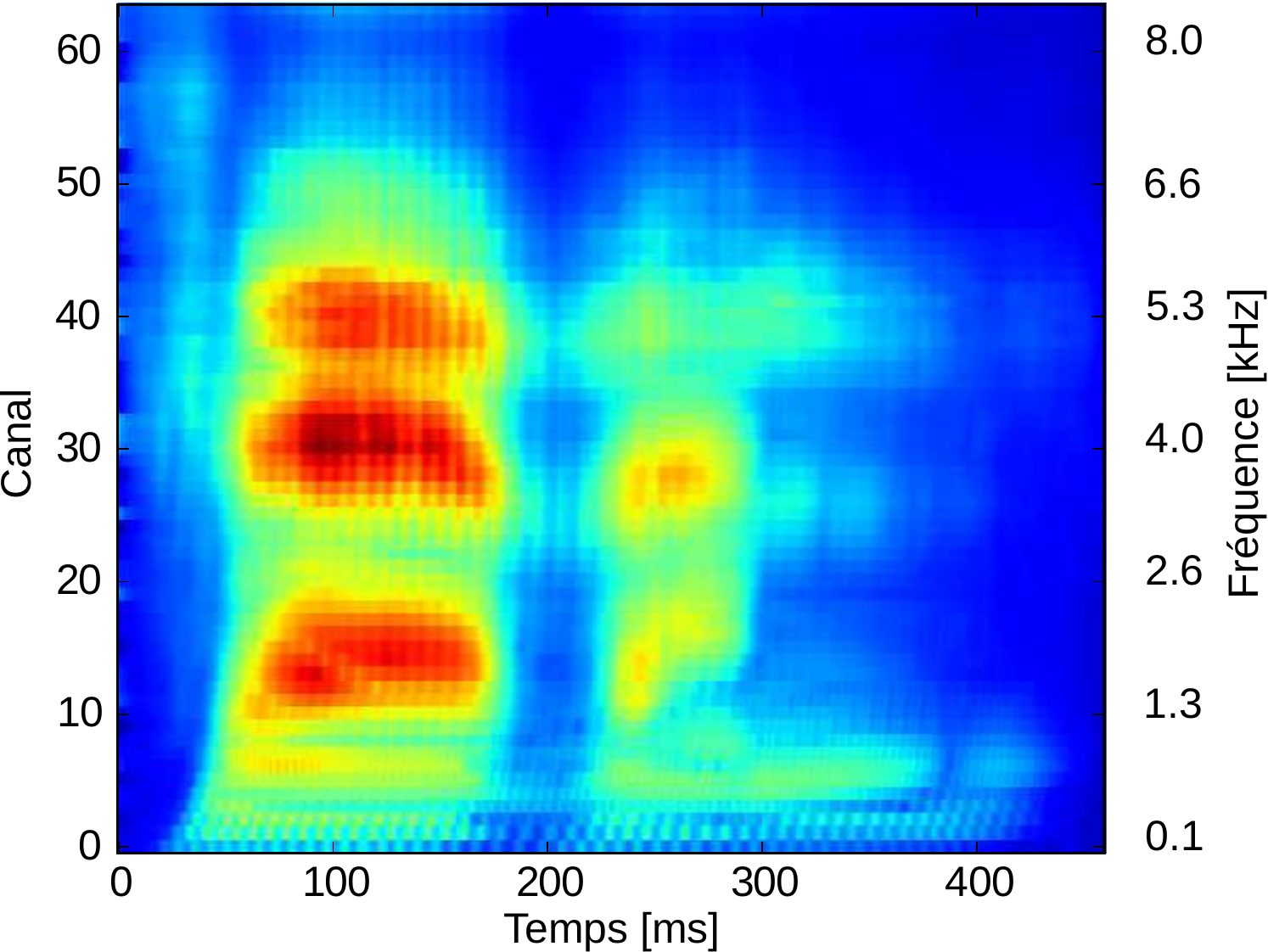}
\caption{Le cochléogramme est une représentation spectro-temporelle du signal acoustique. L'exemple correspond ici à la prononciation du mot anglais \textit{seven} par un locuteur homme. Chaque canal correspond à la sortie d'un filtre cochléaire passe-bande et expose les caractéristiques de modulation temporelle propre à une bande de fréquence. Les paramètres des filtres sont dérivés d'observations neurophysiologiques de la cochlée et d'observations psychoacoustiques, favorisant une plus grande résolution temporelle que spectrale.}
\label{fig:cochleogram}
\end{figure}

Pour ce travail, l'accent est mis sur l'obtention de représentations objets à partir de caractéristiques dérivées des enveloppes des signaux cochléaires.
Ceci permet de capturer de façon générique les formes les plus courantes de modulation en amplitude (AM) et en fréquence (FM), ainsi que les transitoires et les configurations de formants.
La représentation d'entrée aux systèmes qui sont étudiés est le cochléogramme (illustré à la figure~\ref{fig:cochleogram}, page~\pageref{fig:cochleogram}). On cherchera à exploiter l'information contenue dans les différents patrons de modulation spectro-temporelle locale pour caractériser les objets sonores. Dans la suite de l'article, un traitement par blocs/fenêtres et la prise en compte de l'aspect spatial référeront au fait que le cochléogramme sera interprété comme étant une image, par sa nature bidimensionnelle (i.e. représentation temps-fréquence).

\subsection{Projection hiérarchique par dictionnaire}

La formulation mathématique de la projection hiérarchique est la suivante:
soit $\mathbf{S}^{(h)}$ un ensemble de $n$ signaux de dimension $N$, en entrée au niveau hiérarchique $h$, i.e.   $\mathbf{S}^{(h)} = \left[ \mathbf{s}_1 \ldots \mathbf{s}_n \right] \in \Re^{N \times n}$. Soit $\mathbf{D}^{(h)}$ un dictionnaire de $K$ bases de dimension $N$, au niveau hiérarchique $h$, i.e.   $\mathbf{D}^{(h)} = \left[ \mathbf{d}_1 \ldots \mathbf{d}_K \right] \in \Re^{N \times K}$.
La dimension $N$ peut varier selon le niveau de la hiérarchie pour effectuer une réduction de dimension, ou une expansion pour obtenir une représentation surcomplète. La projection des coefficients d'entrée de l'étage inférieur sur le dictionnaire produit un nouvel ensemble de coefficients $\mathbf{C}^{(h)} = \left[ \mathbf{c}_1 \ldots \mathbf{c}_n \right] \in \Re^{K \times n}$, comme montré à l'équation (\ref{eq:coefficients}) et valide pour $h>0$. La pseudo-inverse généralisée (Moore–Penrose) permet d'approximer $\left(\mathbf{D}^{(h)} \right) ^{-1}$, car $\mathbf{D}^{(h)}$ est une matrice rectangulaire. Pour éviter les problèmes d'instabilité numérique, la pseudo-inverse $\left(  \mathbf{D}^{(h)} \right) ^{+}$ sera toutefois calculée par décomposition en valeurs singulières~\cite{RadhakrishnaRao1967}, plutôt que la forme directe d'optimisation au sens des moindres carrés. La transposée du dictionnaire est définie par $\mathbf{D}^{(h) ^T}$.

\begin{align}
\mathbf{C}^{(h)} &= \left(  \mathbf{D}^{(h)} \right) ^{+} \cdot \mathbf{S}^{(h-1)} = \left(  \mathbf{D}^{(h)^T} \mathbf{D}^{(h)} \right) ^{-1} \mathbf{D}^{(h) ^T} \cdot \mathbf{S}^{(h-1)} \nonumber \\
&\text{pour tout niveau abstrait, où} \; h > 0
\label{eq:coefficients}
\end{align}

La projection hiérarchique possède un aspect spatial, où les fenêtres adjacentes sont concaténées et le vecteur résultant projeté sur l'étage supérieur. La reformulation selon les coordonnées $(i,j)$ dans le référentiel local pour chaque étage est donnée à l'équation (\ref{eq:decomposition_hierarchy}). Au premier niveau $h=0$, les signaux d'entrée correspondent à des fenêtres $\mathbf{W}^{(i,j)} \in \Re^{L_C \times L_T}$ distribuées spatialement sur la représentation spectro-temporelle, et converties sous forme de vecteurs colonnes $\mathbf{X}^{(i,j)} \in \Re^{L_C \cdot L_T}$. Les constantes $L_C$ et $L_T$ correspondent respectivement au nombre de canaux et d'échantillons temporels couverts par chaque fenêtre.
Pour tout niveau $h>0$, la projection est appliquée sur la concaténation des coefficients des projections du niveau inférieur $h-1$. Les constantes $M_{(h)}$ et $N_{(h)}$ correspondent au nombre de fenêtres de projection adjacentes considérées respectivement sur l'axe fréquentiel et temporel. 

\newcommand{\concat}{\mathop{\Huge \mathlarger{\mathlarger{\parallel}}}}

\begin{equation}
\mathbf{C}^{(i,j)}_{(h)} = 
\begin{cases}
\mathbf{D}_{(0)}^{+} \cdot \mathbf{X}^{(i,j)}
 & \text{si $h=0$},\\
\mathbf{D}_{(h)}^{+} \cdot \concat_{i=0}^{M_{(h)}-1} \concat_{j=0}^{N_{(h)}-1} \mathbf{C}^{(i,j)}_{(h-1)}
& \text{si $h>0$}. \\
\end{cases}
\label{eq:decomposition_hierarchy}
\end{equation}

Le symbole $\concat$ définit l'opération de concaténation sur $P$ matrices de coefficients adjacentes:
\begin{equation}
\concat_{i=0}^{P-1} \mathbf{C}^{(i)} = 
\begin{bmatrix}
 \mathbf{C}^{(i)} & \mathbf{C}^{(i+1)} & \ldots & \mathbf{C}^{(i+P-1)} 
\end{bmatrix}
\end{equation}

\begin{figure}[htb!]
\centering
\includegraphics[width=1.0\linewidth]{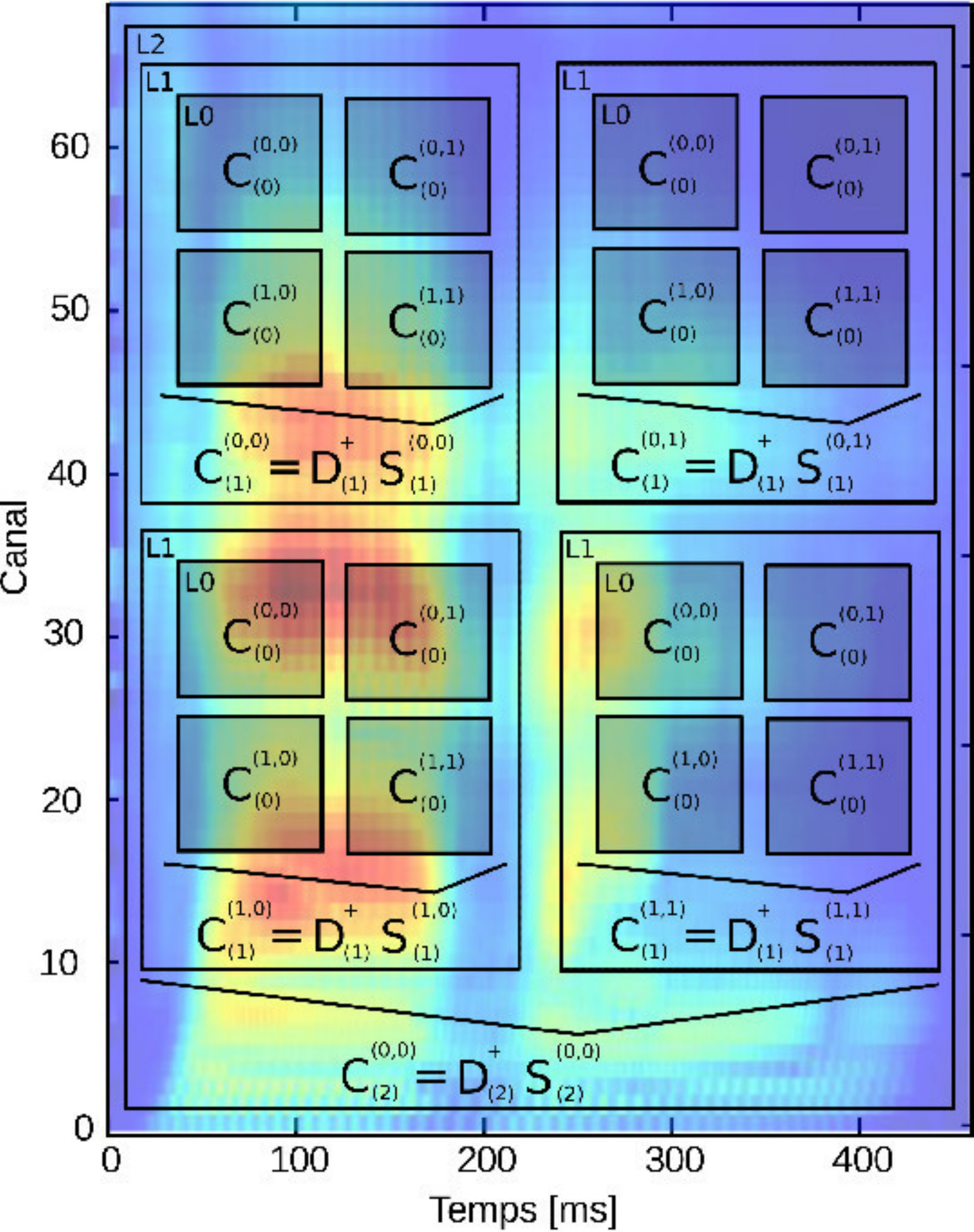}
\caption{Projection hiérarchique par projection successive sur des dictionnaires. L'aspect spatial est considéré par la concaténation des projections adjacentes du niveau inférieur, visant à extraire des caractéristiques couvrant de plus en plus de contexte fréquentiel et temporel. Dans l'exemple donné, $M_{(h)} = N_{(h)} = 2 \; \text{pour} \; h = \lbrace 1, 2 \rbrace$, soit la concaténation de 4 fenêtres adjacentes à chaque niveau.
Les contours figuratifs des fenêtres sont montrés sur le cochléogramme. Il est à noter qu'en réalité, les niveaux L1 et L2 ne couvriront jamais plus que l'aire définie par le niveau L0 (en ombragé).}
\label{fig:context_coord}
\end{figure}

La figure \ref{fig:context_coord} montre plus intuitivement l'aspect spatial de cette projection. Par souci de simplicité, le chevauchement entre les fenêtres n'est pas illustré. 

La projection hiérarchique permet aussi d'allouer un plus grand contexte temporel ou spectral selon le fenêtrage et le chevauchement choisi. Pour éviter les discontinuités causées par le fenêtrage initial $\mathbf{W}^{(i,j)}$ au niveau du cochléogramme, il est possible d'introduire un chevauchement temporel et spectral. Ceci permet de mieux couvrir les caractéristiques du signal avec un nombre limité de bases, au profit d'une représentation de sortie contenant plus de coefficients et où certaines dimensions peuvent devenir fortement corrélées.

\subsection{Analyse en composantes indépendantes}

La qualité des bases obtenues par l'algorithme d'apprentissage non-supervisé du dictionnaire $\mathbf{D}^{(h)}$  pour chacun des étages $h$ est ultimement mesurée par l'effet sur la performance par exemple d'un système de classification. Toutefois, l'interprétation visuelle des bases (i.e. caractéristiques spatio-temporelles extraites) est pourtant importante pour valider que la projection hiérarchique permet bien l'extraction de structures complexes dans le signal. L'analyse en composantes indépendantes (ICA)~\cite{Hyvarinen2000} est une méthode de décomposition linéaire (voir Équation \ref{eq:coefficients}) favorisant une représentation par objets d'un signal en forçant l'indépendance entre les composantes des objets. Cette décomposition définit des bases en se basant donc sur un critère de maximisation de leurs indépendances statistiques. Comparativement à l'analyse en composantes principales (PCA), il s'agit d'un critère plus strict que la décorrélation, car les moments d'ordres supérieurs à 2 sont considérés. Pour cette raison, il y a contrainte que les composantes indépendantes doivent posséder une distribution non-gaussienne. 

L'implémentation FastICA~\cite{Hyvarinen1999a} a été utilisée pour les expériences. L'avantage principal est qu'il n'y a pas de méta-paramètres à choisir, comme une constante de régularisation définissant le compromis entre la parcimonie (coefficients ou bases) et l'erreur de reconstruction. Seule une fonction de contraste pour l'approximation de la néguentropie est requise, mais dont le choix n'est important que pour optimiser la performance de l'algorithme selon le type de non-gaussianité des composantes~\cite{Hyvarinen1999a}.
En fait, tant qu'il y a critère de minimisation de l'information mutuelle, les caractéristiques extraites seront localisées en temps et en fréquence si l'apprentissage est effectué sur des sons naturels ou des signaux de parole~\cite{lewicki2002,lee2002}.

Le problème d'estimation des sources par ICA devient plus complexe si le nombre de composantes indépendantes est supérieur au nombre de mixtures observées, car le processus de mélange est non-inversible en raison d'une perte d'information~\cite{hyvarinen2001}. Il s'agit de la situation où la représentation est sur-complète. Pour la présente architecture, chaque étage $h$ de la hiérarchie produit un vecteur de coefficients dont la dimension est inférieure à celle de la sortie de l'étage précédent. Ceci est principalement dû à la concaténation des projections adjacentes. Il s'agit alors du cas de représentation sous-complète. 
Un des avantages de cette situation est qu'en considérant l'aspect de parcimonie, il y a capacité inhérente à classifier les signaux d'entrée~\cite{Wang2008a}.  
 Pour une tâche d'extraction de caractéristiques, il est alors possible de compresser l'information sans explosion du nombre de dimensions. Alternativement, une réduction de dimension des vecteurs d'entrée est recommandée dans le cas de représentation sous-complète~\cite{Naik2011}, mais l'utilisation de l'analyse par composantes principales (PCA) reste mitigée. Sachant l'ambiguïté de ICA au niveau de la variance des composantes (i.e. assumée fixe à 1), les composantes indépendantes ne sont pas forcément contraintes dans le sous-espace défini par PCA~\cite{Porrill1998}.
La projection hiérarchique proposée vise avant tout à décomposer le signal en composantes parcimonieuses, mais sans la contrainte de devoir reconstruire le signal. Les différents problèmes énoncés précédemment sont alors de moindre importance comparativement à une application en codage de la parole.


\vspace{1em}

\section{Expérience 1: Recherche non-supervisée de bases}

L'apprentissage non-supervisé permet d'obtenir une représentation naturellement adaptée au signal à modéliser ou reconnaître. Peu d'effort manuel est alors requis dans le choix des paramètres optimaux du système. 
Dans cette expérience, les types de bases obtenues lorsque l'apprentissage est effectué sur des catégories de sons différentes (e.g. parole et musique) ont été comparés. La base de données TIMIT~\cite{TIMIT1993} consistant en 330 minutes de parole continue (sous forme de phrases) a été utilisée pour la catégorie de parole. Pour la catégorie de musique, 73 minutes de musique classique orchestrale~\cite{Perlman2002} ont été utilisées. Enfin, pour la catégorie des sons naturels, 60 minutes d'enregistrement d'une plage tropicale~\cite{NatureSoundSeries2007} et 60 minutes d'enregistrement d'une forêt de montagne~\cite{NatureSoundSeries2006} ont composé la base de données d'apprentissage spécifique.

\subsection{Description du système}

Un filtre de pré-accentuation découlant de la réponse spectrale observée physiologiquement chez l'humain~\citep{Huber2001} est d'abord utilisé pour rehausser les moyennes fréquences.
Une analyse spectro-temporelle est ensuite effectuée par un banc de $64$ filtres Gammatone~\cite{Hohmann2002}. Le banc de filtres a été corrigé pour le décalage de phase entre les canaux, ce qui fait qu'une impulsion glottale produira une réponse instantanée sur tous les canaux. La plage de fréquence couverte est de $[0, 8000 \;\text{Hz}]$ et l'espacement des filtres est linéaire sur l'échelle de Mel~\citep{Stevens1937}. La largeur de bande des filtres est grande pour privilégier la résolution temporelle à la sortie des canaux, qui favorisera une meilleure modélisation des transitoires tout en évitant la résolution de chacune des harmoniques lors des segments voisés.
L'étape suivante dans le calcul du cochléogramme est une rectification simple-alternance suivie d'une compression par une racine cubique~\citep{Avendano2004}. L'usage d'une compression fortement non-linéaire offre l'avantage d'augmenter radicalement le contraste entre les canaux de faibles amplitudes. Les modulations d'amplitude en moyenne et haute-fréquence (où l'énergie est souvent moindre) deviennent alors plus comparables à celles en basse-fréquence. Finalement, un filtre Butterworth passe-bas d'ordre 1, avec une fréquence de coupure à $40$ Hz, permet le lissage du spectre et réduit ainsi l'effet des impulsions glottales sur les caractéristiques de modulation d'amplitude extraites par l'algorithme d'analyse en composante indépendante (ICA).

\begin{figure}[htb!]
\centering
\includegraphics[width=1.0\linewidth]{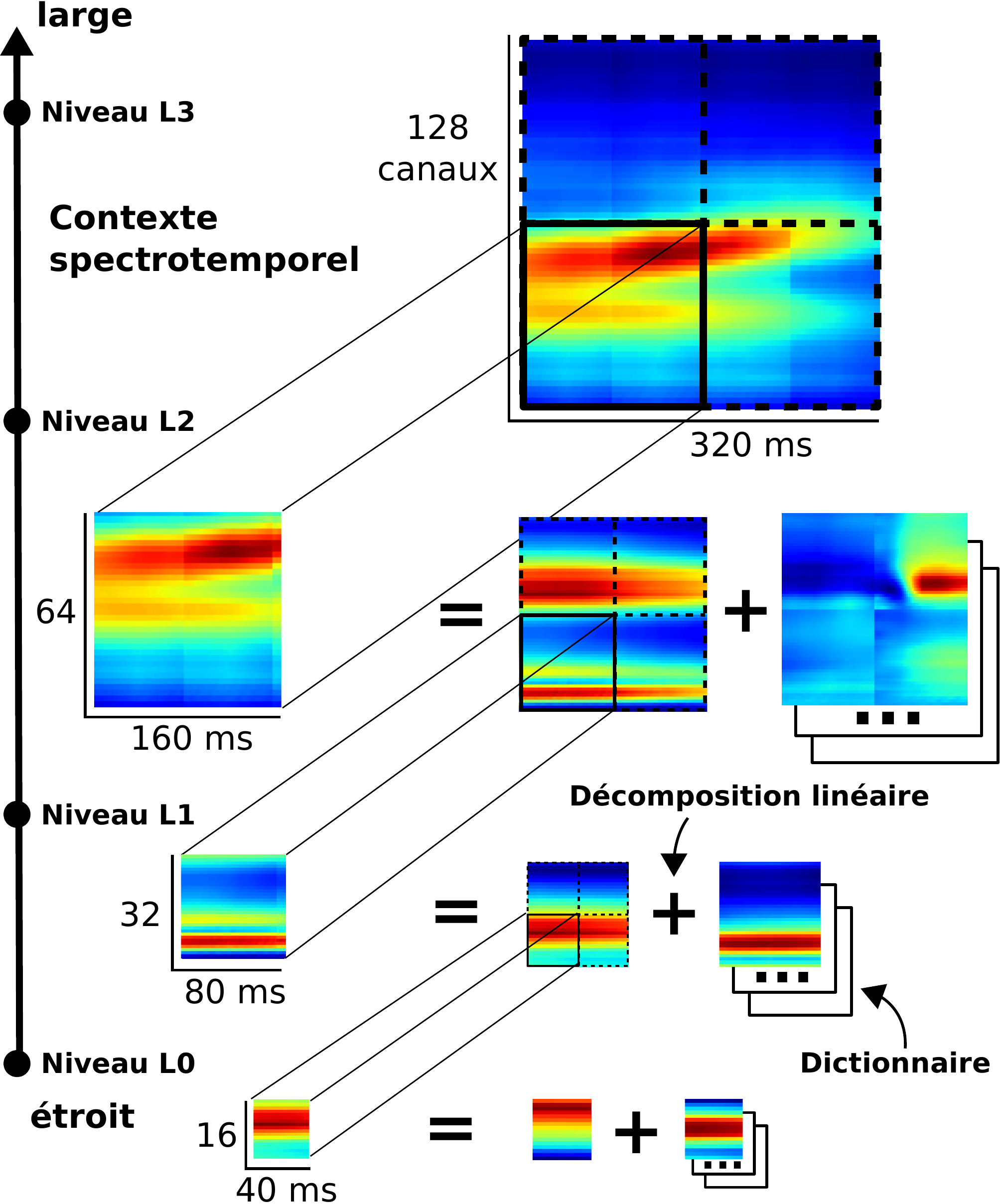}
\caption{Projection hiérarchique jointe dans l'axe temporel et fréquentiel pour des dictionnaires (ensembles de bases) de niveau L0, L1, L2 et L3. Par exemple, le dictionnaire L2 est basé sur les projections du niveau inférieur L1, lui-même basé sur les projections du niveau L0. Une grille spatiale de taille 2x2 est ici utilisée pour joindre les coefficients des 4 blocs de projection adjacents. Chaque bloc de projection est une composition linéaire (sommation ou négation) des prototypes définis dans le dictionnaire spécifique à l'étage. Avec cette méthode de projection, le dictionnaire finira par définir des objets complexes et analogues à des entités syllabiques couvrant un contexte spectro-temporel de plus en plus large.}
\label{fig:projection}
\end{figure}

Une intégration du contexte spatio-temporel permet d'extraire des caractéristiques de plus en plus complexes. Sans poser d'hypothèses sur la nature des composantes des objets, mais en augmentant simplement le contexte spatial et temporel, des structures cohérentes et pertinentes sont apprises par les bases. Elles représentent des parties élémentaires d'objets sonores. La figure \ref{fig:projection} montre en quoi une projection hiérarchique (sans chevauchement) exploite l'augmentation du niveau d'abstraction, ce qui permet la considération d'un plus large contexte spatio-temporel dans la représentation d'un signal de parole.

Pour cette expérience, les paramètres utilisés pour effectuer la projection hiérarchique sont les suivants:
Le nombre d'étages de projection égal à 3, la taille $K$ du dictionnaire à chaque étage étant respectivement de 128, 256, et 256.
Il y a un fenêtrage initial $\mathbf{W}^{(i,j)}$ de $L_C = 16$ canaux par $L_T = 40$ ms, sans chevauchement. Le cochléogramme comporte 64 canaux avec une fréquence d'échantillonnage de 1000Hz. Il y a concaténation de $M_{(h)} = 2$ blocs au niveau spectral, et $N_{(h)} = 3$ bloc au niveau temporel. Cette configuration fera en sorte que le dernier niveau de projection couvrira $64$ canaux par $360$ ms. L'apprentissage non-supervisé du dictionnaire est effectué par analyse en composantes indépendantes (ICA). Les étages L0 et L1 sont entraînés avec 100,000 vecteurs de coefficients, tandis que pour l'étage L2, le nombre est limité entre 20,000 et 50,000 vecteurs d'apprentissage. Connaissant le grand contexte temporel des bases de haut-niveau, le nombre d'exemples pour l'entraînement est ultimement limité par la taille spécifique de la base de données.

L'entraînement des dictionnaires selon les étages de la hiérarchie est effectué de façon itérative, donc successivement du niveau L0 jusqu'au niveau L2. Ceci montre les contraintes de dépendance avec les étages inférieurs.

\subsection{Résultats}

\begin{figure}[htb!]
\centering
\subfigure[Parole isolée (TI46)]{
\includegraphics[width=1.0\linewidth]{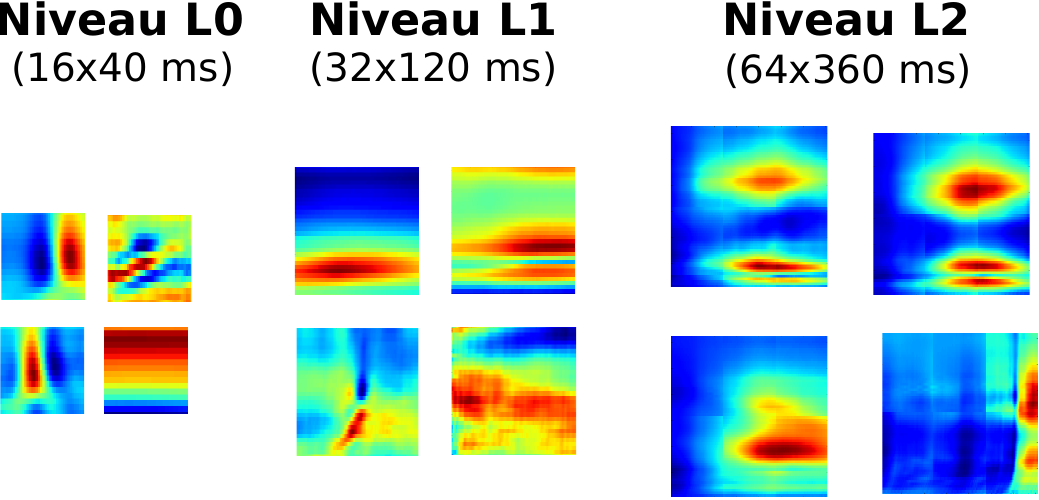}
\label{fig:bases_speech}
}

\subfigure[Musique classique (Vivaldi)]{
\includegraphics[width=1.0\linewidth]{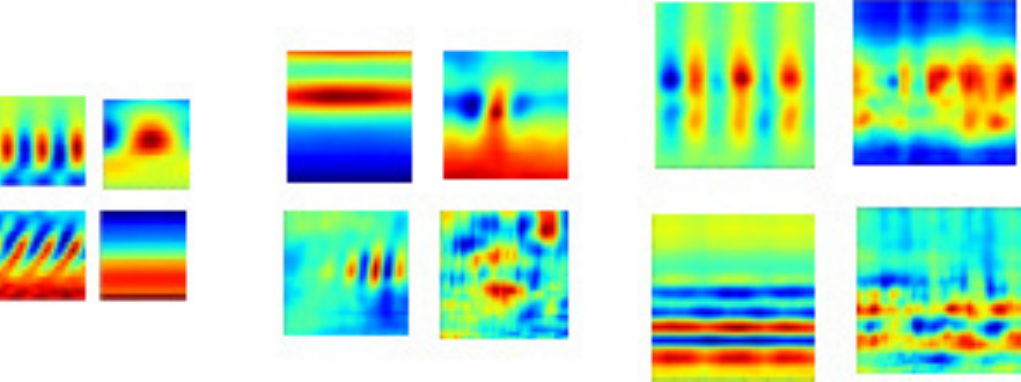}
\label{fig:bases_music}
}

\subfigure[Nature (forêt tropicale et plage)]{
\includegraphics[width=1.0\linewidth]{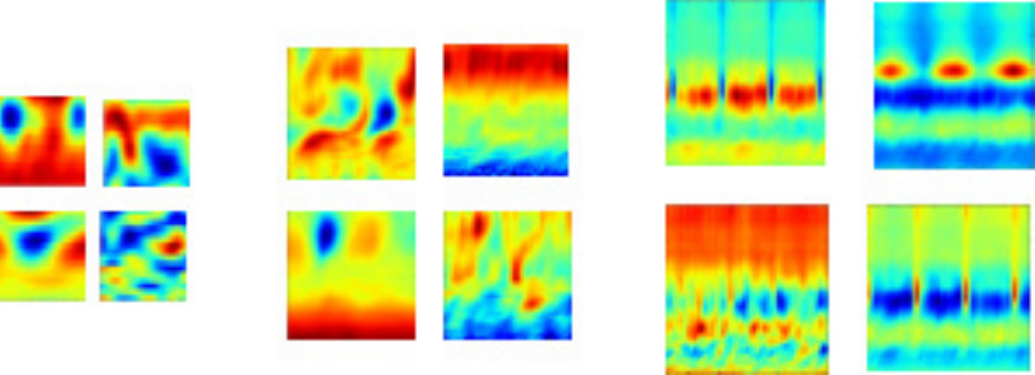}
\label{fig:bases_nature}
}

\caption{Bases apprises pour différents types de sons naturels: \subref{fig:bases_speech} parole isolée, \subref{fig:bases_music} musique classique et \subref{fig:bases_nature} sons de nature. On remarque les différences notables des bases apprises entre les catégories de sons, peu importe le niveau dans la projection hiérarchique. Les bases de haut-niveau apprises pour la parole font ressortir des configurations de formants, alors que pour la musique, il s'agit plutôt d'harmoniques constantes ou de séquences de tonalités brèves. Pour les sons de nature, les bases sont aussi très variées.}
\label{fig:bases}
\end{figure}

Les bases obtenues après apprentissage modélisent bien les caractéristiques propres aux différentes catégories de sons, comme il est illustré à la figure \ref{fig:bases} ( page~\pageref{fig:bases}). Peu importe le type de sons, il y a représentation parcimonieuse du signal à chaque étage (non montré). Cette propriété est évaluée objectivement par la mesure de kurtosis (moment centré d'ordre 4) du vecteur de coefficients. Peu importe le type de sons, il y a aussi représentation par objets du signal, comme révélée par une inspection visuelle et subjective des bases. Avec plus de données d'apprentissage pour chacune des catégories, des bases supplémentaires de plus haut-niveau (i.e. plusieurs secondes) auraient pu définir des patrons de modulation spectro-temporelle encore plus complexe.

\section{Expérience 2: reconnaissance de mots isolés}

Pour un système de reconnaissance de forme, il devrait idéalement y avoir un passage graduel de l'apprentissage non-supervisé à un apprentissage supervisé. Alors qu'une certaine ambiguïté sur la nature du signal est permise pour l'étage d'extraction de caractéristiques, une décision stricte devra être prise à l'étage de classification. Dans cette expérience, une représentation dérivée de la projection hiérarchique est utilisée pour construire un système de reconnaissance de mots isolés basé sur une modélisation statistique par modèle de Markov caché (HMM). 
La base de données TI46~\citep{Liberman1993} est couramment utilisée pour tester les systèmes automatiques de reconnaissance de type mots isolés et dépendant du locuteur. La base de données contient seulement 46 classes (alphabet, chiffres et commandes). On dispose de 16 locuteurs (8 hommes et 8 femmes) et de 26 prononciations par locuteur par mot. Pour les ensembles d'entraînement et de test, on dispose respectivement de 10 et 16 prononciations par locuteur. Le partitionnement standard pour cette base de données a été utilisé. Pour tester la capacité de généralisation, le système de reconnaissance est entraîné de façon à être indépendant du locuteur.

Le HMM est un modèle statistique générateur d'une séquence d'observations basé sur un espace d'états et une topologie de transitions~\cite{rabiner1989}, communément utilisé en reconnaissance de la parole. Un modèle de mot entier et une structure gauche-droite à 16 états ont été utilisés pour chacune des classes. Lors de l'évaluation d'un signal acoustique contenant un mot (mais dont l'identité reste à déterminer), le modèle ayant le maximum de vraisemblance à avoir généré la séquence d'observations est retenu comme le mot prononcé le plus probable.

\subsection{Description du système proposé}

\begin{figure}[htb!]
\centering
\includegraphics[width=0.85\linewidth]{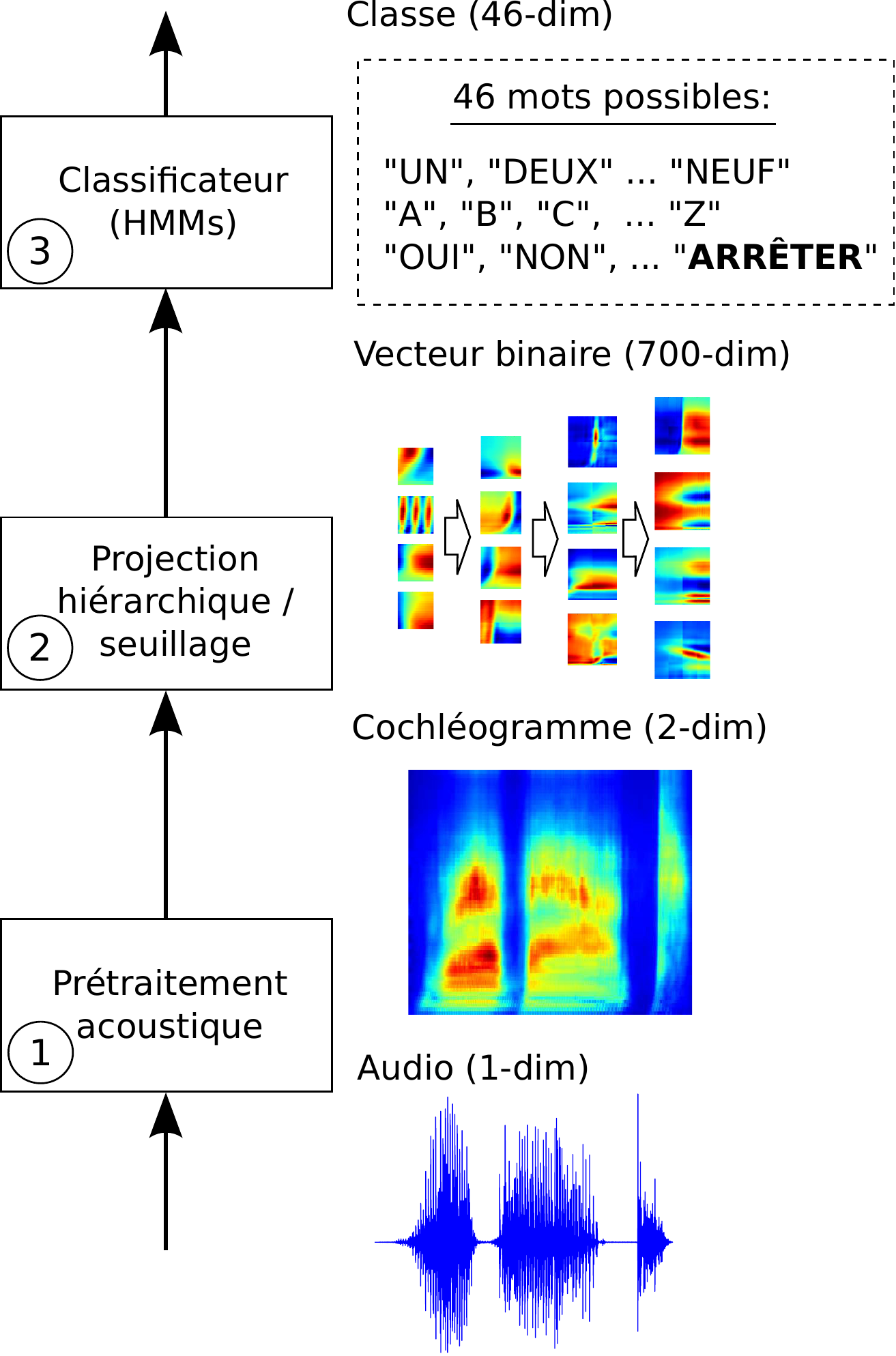}
\caption{Étapes de traitement dans l'architecture proposée, soit le prétraitement acoustique, la projection hiérarchique et la classification. La dimension des représentations augmente jusqu'à la sortie de l'étage de classification, où l'identité du mot dans le signal sonore est déterminée.}
\label{fig:global_architecture}
\end{figure}

Les différents étages composant l'architecture proposée sont illustrés à la figure~\ref{fig:global_architecture} (page~\pageref{fig:global_architecture}), qui met l'emphase sur l'évolution de la dimension des représentations selon les étages. Il y a explosion des dimensions à la sortie d'une projection hiérarchique, mais la parcimonie de la représentation assure une activation restreinte à certaines dimensions seulement. Ceci permettra d'augmenter la robustesse au bruit du système, considérant de plus que les bases auront été adaptées aux caractéristiques de la parole.

\begin{figure*}[htb!]
\centering
\includegraphics[width=0.85\linewidth]{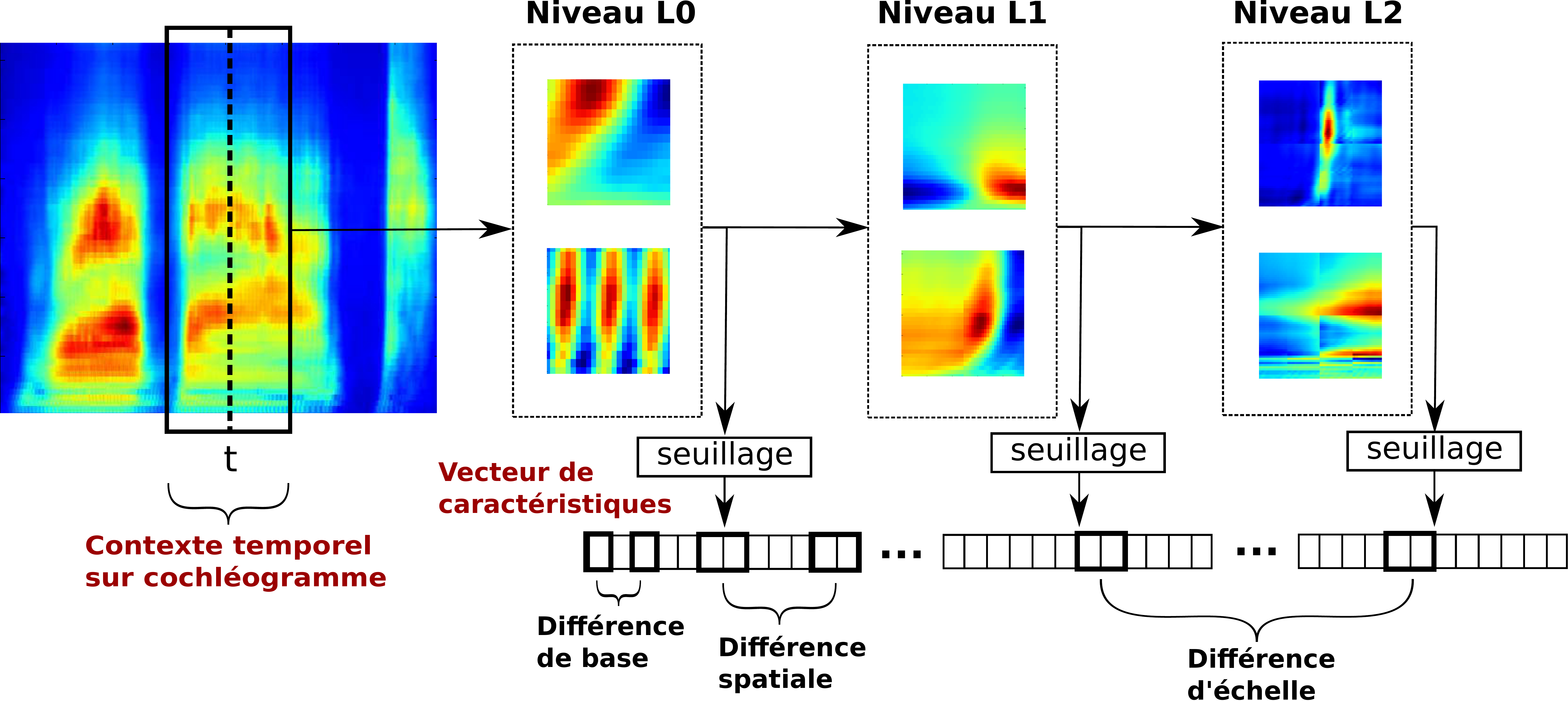}
\caption{Création du vecteur de caractéristiques à partir de la projection hiérarchique. À partir du cochléogramme (représentation temps-fréquence), il y a création d'un vecteur colonne représentant les caractéristiques de différentes bases, sur plusieurs positions spatiales et échelles spatio-temporelles d'analyse. Il est possible d'appliquer un traitement spécifique sur les représentations de sortie de chacun des étages, visant à favoriser la parcimonie et la production de coefficients binaires par des mécanismes de seuillage et de compétition.}
\label{fig:feature_vector}
\end{figure*}

Il doit être possible à chaque intervalle de temps $\frac{1}{F_{sous}}$ dans le signal de parole de définir un vecteur de caractéristiques multi-échelle pour la modélisation statistique (e.g. modèle de Markov caché). La constante $F_{sous}$ est la fréquence d'échantillonnage du vecteur de caractéristiques correspondant à une observation discrète. La séquence d'observations produite sert ensuite d'entrée à l'étage de classification. La figure \ref{fig:feature_vector} (page~\pageref{fig:feature_vector}) illustre comment cette opération est effectuée pour produire un vecteur binaire, parcimonieux et à grande dimension. La projection hiérarchique est efficace en termes de calculs dans le sens où des techniques de programmation dynamique permettent d'éviter le recalcul  d'une même projection (bloc ou fenêtre) si les mêmes coefficients sont utilisés par plusieurs projections d'ordre supérieur.

L'expression générale d'une modélisation acoustique par modèle de mixture est donnée à l'équation (\ref{eq:mixture_prob}).
La densité de probabilité $p \left(\mathbf{x} \right)$ découle de la contribution de $M$ mixtures, décrites par une probabilité a priori $p \left( i \right)$ et une vraisemblance $p \left(\mathbf{x} \vert i \right)$. 

\begin{align}
& p \left(\mathbf{x} \right) = \sum_{i=0}^{M-1} p \left( i \right)  p \left(\mathbf{x} \vert i \right)
\label{eq:mixture_prob} \\
& \text{avec les contraintes \hspace{0.5cm}}  \sum_{i=0}^{M-1} p \left( i \right) = 1  \text{\hspace{0.5cm} et \hspace{0.5cm}} p \left( i \right)  \in \left[ 0,1 \right] \nonumber
\end{align}

\newcommand{\E}{\mathrm{E}}

\newpage
Avec l'utilisation de représentations binaires parcimonieuses, les mixtures de gaussiennes (GMMs) ne sont plus adaptées pour la modélisation acoustique dans une architecture HMM, car elles modélisent des variables aléatoires continues et non discrètes. Le cas d'une mixture de fonctions gaussiennes multivariables à n-dimensions est défini à l'équation (\ref{eq:gaussian_prob}). Les moments de premier ordre $\mathbf{\mu}_i = \E \left( X \right)$ et de deuxième ordre $\mathbf{\Sigma}_i = \text{Cov} \left( X \right)$ paramétrisent chacune des $M$ mixtures.
\begin{align}
p \left(\mathbf{x} \vert i \right) &= \frac{1}{{\left( 2 \pi \right)}^{\frac{n}{2}} {\vert \mathbf{\Sigma}_i \vert}^{\frac{1}{2}}} e^{- \frac{1}{2} (\mathbf{x} - \mathbf{\mu}_i)^T \mathbf{\Sigma}^{-1}_i (\mathbf{x} - \mathbf{\mu}_i)}
\label{eq:gaussian_prob}
\end{align}

Les mixtures de Bernoulli (BMMs) offrent une alternative appropriée pour construire un modèle générateur d'observations binaires. Cette divergence par rapport aux systèmes conventionnels de reconnaissance de la parole permet de définir un espace de paramètre, au niveau du modèle statistique, qui disposera des mêmes propriétés que la représentation d'entrée: les paramètres des mixtures seront de grande dimension et parcimonieux.

Prenons le cas spécial d'une distribution de Bernoulli de dimension $N$, où les dimensions sont considérées comme statistiquement indépendantes. La fonction de probabilité est définie à l'équation (\ref{eq:bernoulli_prob}), où $x_n \in \lbrace 0, 1 \rbrace$ est l'élément à la dimension $n$ du vecteur binaire $\mathbf{x}$ de dimension $N$, et $p_{i,n} \in \left[ 0,1 \right]$ est le paramètre de la dimension $n$ associé à la mixture $i$. Ce dernier reflète intuitivement la probabilité moyenne d'avoir une activation positive (i.e. 1) sur une certaine dimension.

\begin{equation}
p \left(\mathbf{x} \vert i \right) = \prod_{n=1}^{N} p_{i,n}^{x_n} \left( 1 - p_{i,n} \right) ^ {1 - x_n}
\label{eq:bernoulli_prob}
\end{equation}

L'entraînement des paramètres des mixtures et du modèle markovien (e.g. probabilités de transitions) est fait par l'algorithme Expectation-Maximisation (EM). Comme les paramètres $p_{i,n}$ correspondent à la probabilité moyenne d'apparition sur chacune des dimensions, la mise à jour est identique à celle utilisée pour le paramètre de moyenne des mixtures de gaussiennes. Pour un processus de Bernoulli multivariable, on fait l'hypothèse d'indépendance statistique de chacune des dimensions. Dans le cas où les caractéristiques d'entrée découlent d'une analyse par composantes indépendantes (ICA), cette hypothèse est réaliste.

L'utilisation de mixtures de Bernoulli est relativement rare en reconnaissance de forme, et se concentre sur les travaux de quelques chercheurs seulement. Les mixtures de Bernoulli ont d'abord été investiguées pour la reconnaissance d'images binaires~\cite{Juan2004a, Romero2007, Ad2009}. L'effet des paramètres initiaux sur la convergence a ensuite été étudié~\cite{Juan2004}. Finalement, l'intégration avec un HMM a été effectuée~\cite{Ad2009, Gimenez2009}, mais toujours en reconnaissance d'image et non de parole. 

Les avantages des mixtures de Bernoulli dans l'application présente sont nombreux. Premièrement, les paramètres d'une mixture de Bernoulli définissent essentiellement des prototypes dont la visualisation/interprétation est facilement concevable (surtout dans le cas du traitement d'image~\cite[e.g.][]{Juan2004a}). Deuxièmement, un processus de Bernoulli considère seulement la moyenne, et ignore la variance.  Il y a donc une meilleure convergence, car nul besoin de techniques de seuillage des paramètres des mixtures pour éviter la surspécialisation (i.e. variance nulle ou concentration autour d'une seule observation). L'initialisation des paramètres est aussi moins susceptible à porter problème dans le cas des mixtures de Bernoulli, où quelques techniques simples permettent d'éviter les configurations pathologiques de paramètres~\cite{Juan2004}. Enfin, la parcimonie dans l'espace des paramètres peut être forcée si le nombre de dimensions ou variables latentes est large, dans le cas d'un modèle générateur par variables latentes binaires~\cite{Henniges2010}. Il y a donc un potentiel intéressant pour conserver la similitude avec les représentations d'entrée, qui sont parcimonieuses et à haute-dimensionnalité.

Pour l'extraction des caractéristiques, les paramètres sont légèrement différents de l'expérience précédente portant sur l'émergence de bases non-supervisées pour diverses catégories de sons.
La taille $K$ du dictionnaire pour chacun des 3 étages est respectivement de 64, 128, et 256. L'entraînement utilise maximalement 25,000 exemples de vecteurs de coefficient lors de l'apprentissage non-supervisé des bases avec l'algorithme d'analyse en composantes indépendantes (ICA).
La fréquence d'échantillonnage $F_{sous}$ du vecteur de caractéristiques est de $100$ Hz. Il y a un fenêtrage initial $\mathbf{W}^{(i,j)}$ de $L_C = 32$ canaux par $L_T = 40$ ms, avec $50\%$ de chevauchement temporel et spectral. Il y a concaténation de $M_{(h)} = 2$ blocs au niveau spectral, et $N_{(h)} = 2$ bloc au niveau temporel, avec un chevauchement de $25\%$ au niveau des blocs abstraits (i.e. pour $h>0$). Cette configuration fera en sorte que le dernier niveau de projection couvrira $64$ canaux par $160$ ms, donc de l'ordre de grandeur suprasegmental ou syllabique.
Pour l'apprentissage du modèle acoustique par mixture de Bernoulli à 8 composantes, 50 itérations de l'algorithme EM ont été effectuées.
Chaque HMM possède une topologie gauche-droite à 16 états. L'influence de chaque méta-paramètre sur les taux de reconnaissance a été validée pour obtenir cette configuration optimale.

\subsection{Description du système de référence}

L'étage d'extraction des caractéristiques le plus commun en reconnaissance de parole et basé sur les coefficients cepstraux sur l'échelle fréquentielle de Mel (MFCC) et de leurs dérivées temporelles. Les paramètres utilisés ont été tirés du livre de référence de la suite HTK~\cite{Young1997}.
Après un filtre de préaccentuation ($\alpha = 0.97$) et un fenêtrage par Hamming du signal d'entrée, une transformée de Fourier à court terme est appliquée. Les amplitudes spectrales sont alors projetées sur l'échelle de Mel, consistant en des filtres triangulaires chevauchants. Les logarithmes des amplitudes sont alors pris à la sortie du banc de filtres, puis une transformée en cosinus discrète est appliquée pour obtenir les coefficients cepstraux. Seuls les premiers coefficients sont considérés, car ils portent l'information grossière du profil spectral. Il y a rehaussement standard et normalisation de la moyenne des coefficients cepstraux, puis calcul des dérivés premières (delta) et secondes (delta-delta). Le vecteur de caractéristiques possède en tout 39 dimensions: 1 log-énergie,  12 coefficients cepstraux, 13 delta, 13 delta-delta.
Intuitivement, les MFCCs modélisent donc l'enveloppe spectrale et son évolution locale à chaque fenêtre de temps, par un vecteur dense à faible-dimensionnalité.
Pour l'apprentissage du modèle acoustique par mixture de gaussiennes à 4 composantes, 50 itérations de l'algorithme EM ont été effectuées.
Chaque HMM possède une topologie gauche-droite à 16 états. L'influence de chaque méta-paramètre sur les taux de reconnaissance a été validée pour obtenir cette configuration optimale.

La base de données NOISEX-92~\cite{Varga1993} a été utilisée pour simuler l'effet indésirable d'un bruit additif durant l'entraînement et l'évaluation des systèmes de reconnaissance. Les bruits sont réalistes et non-stationnaires, sauf dans le cas du bruit blanc.
L'entraînement en condition propre est effectué avec les données originales non-bruitées.
Pour l'entraînement multi-condition, chaque fichier de parole de l'ensemble d'entraînement original a été mélangé avec un des bruits utilisés pour l'ensemble de test. Le bruit est choisi aléatoirement pour chacun des fichiers, et le rapport signal-à-bruit (RSB) a été fixé à 20 dB. Il ne s'agit donc que d'un faible niveau de bruit. Toutefois, l'ensemble d'entraînement contient maintenant de l'information sur tous les bruits qui seront présentés avec l'ensemble de test, ce qui devrait réduire la disparité entre l'ensemble d'entraînement et de test. Dans ces conditions, de meilleures performances à bas rapports signal-à-bruit (RSB) sont attendues, car la variabilité induite par le bruit pourra être modélisée durant l'entraînement. L'entraînement d'un système de reconnaissance de parole avec un ensemble d'entraînement bruité permet de voir en quoi le système tire avantage de cette information additionnelle.

Il n'existe pas à notre connaissance de travaux ayant utilisé toutes les 46 classes de la base de données TI46 pour évaluer les taux de performance de reconnaissance. Il est très commun que les sous-ensembles restreints TI-20 (20 classes), TI-ALPHA (26 classes), ou même seulement les chiffres (10 ou 11 classes) soient utilisés. Un avantage est que l'utilisation des 46 classes rend le problème de classification plus difficile, car la confusion entre les classes augmente. Une comparaison directe des performances avec la littérature est donc impossible. Toutefois, les configurations optimales ont été dérivées indépendamment pour le système proposé et le système de référence, ce qui assure une comparaison juste et sans biais. L'objectif est avant tout de démontrer la polyvalence du système proposé, en terme de robustesse, sur différents types d'entraînement (i.e. avec ou sans bruit). Plusieurs techniques d'optimisation (e.g. modèle de silence en début et fin de mot) pourraient être ajoutés aux deux systèmes dans le but d'augmenter les taux de reconnaissance absolue.

\subsection{Résultats}

Le système proposé (SPARSE) et le système de référence (MFCC) ont été évalués sur l'ensemble de données de test bruité après un entraînement en condition propre (voir Tableau \ref{tb:perf_train_clean}). 
On remarque une dégradation des performances de près de $12\%$ (relatif) pour le système MFCC utilisé avec le bruit de conversation, lorsqu'on compare la condition de test propre avec les différents bruits à un rapport signal-à-bruit (RSB) de 40 dB. Le système SPARSE est beaucoup plus robuste dans cette condition, avec une dégradation maximale de $1.2\%$ (relatif). Pour l'intervalle de RSBs entre 10 dB et 40 dB, les performances de ce dernier sont en majorité supérieures au système de référence, sauf dans le cas du bruit blanc gaussien et du bruit de salle de machine. Ces types de bruits sont propices à 
poser problème avec les mécanismes de seuillage utilisés pour produire une représentation binaire. Les caractéristiques décrivant plutôt les zones de faible énergie (e.g. silences ou pause courte), qui étaient ignorées durant l'entraînement en condition propre, peuvent maintenant affecter les autres caractéristiques fiables. Ceci engendre une grande disparité avec le modèle appris, qui ne peut alors plus généraliser correctement. Néanmoins, pour les autres bruits réalistes, l'usage de caractéristiques parcimonieuses et à grandes dimensions est une alternative efficace pour améliorer la robustesse comparativement au système de référence.

\begin{table}[htb!]
\small
\centering
\caption{Taux de reconnaissance avec entraînement en conditions propres et test en conditions adverses, pour différents types de bruits (babble, destroyerengine, volvo, white) et RSBs (-5 dB à 40 dB). Les résultats montrent que le système proposé SPARSE permet une meilleure généralisation à haut RSBs, comparativement au système de référence MFCC. Le bruit blanc gaussien et le bruit de salle de machine semblent toutefois poser problème au système.}
\label{tb:perf_train_clean}

\subtable[Bruit de conversation]{
\begin{tabular}{| r | r | r | r | r | r | r |}
\hline
RSB & -5 dB & 0 dB & 10 dB & 20 dB & 40 dB & Propre\\
\hline \hline
MFCC & \textbf{4.1} & \textbf{6.1} & 16.8 & 36.9 & 70.1 & 78.4  \\
SPARSE & 2.8 & 5.0 & \textbf{19.5} & \textbf{47.6} & \textbf{92.0} & \textbf{93.1} \\
\hline
\end{tabular}
\label{tb:perf_train_clean_babble}
}

\subtable[Bruit de salle de machine d'un contre-torpilleur (bateau)]{
\begin{tabular}{| r | r | r | r | r | r | r |}
\hline
RSB & -5 dB & 0 dB & 10 dB & 20 dB & 40 dB & Propre\\
\hline \hline
MFCC & \textbf{2.5} & \textbf{4.2} & \textbf{14.9} & \textbf{36.9} & 68.3 & 78.4  \\
SPARSE & 2.5 & 3.3 & 8.0 & 21.8 & \textbf{77.6} & \textbf{93.1} \\
\hline
\end{tabular}
\label{tb:perf_train_clean_destroyerengine}
}

\subtable[Bruit intérieur d'une voiture]{
\begin{tabular}{| r | r | r | r | r | r | r |}
\hline
RSB & -5 dB & 0 dB & 10 dB & 20 dB & 40 dB & Propre\\
\hline \hline
MFCC & \textbf{25.6} & \textbf{36.5} & 54.2 & 67.7 & 78.1 & 78.4  \\
SPARSE & 18.2 & 32.6 & \textbf{63.6} & \textbf{84.9} & \textbf{93.2} & \textbf{93.1} \\
\hline
\end{tabular}
\label{tb:perf_train_clean_volvo}
}

\subtable[Bruit blanc gaussien]{
\begin{tabular}{| r | r | r | r | r | r | r |}
\hline
RSB & -5 dB & 0 dB & 10 dB & 20 dB & 40 dB & Propre\\
\hline \hline
MFCC & 2.2 & 2.3 & \textbf{15.2} & \textbf{50.3} & 77.9 & 78.4  \\
SPARSE &  \textbf{2.2} &  \textbf{2.5} & 7.2 & 19.5 & \textbf{90.8} & \textbf{93.1} \\
\hline
\end{tabular}
\label{tb:perf_train_clean_white}
}
\end{table}

Enfin, le système proposé (SPARSE) et le système de référence (MFCC) ont été évalués sur l'ensemble de données de test bruité après un entraînement multi-condition (voir Tableau \ref{tb:perf_train_noisy}). Rappelons que les bruits ont tous été présentés durant l'entraînement à un RSB fixe de 20 dB.
On remarque une dégradation considérable des performances pour le système MFCC, lorsqu'on compare avec l'entraînement en conditions de test propres. Cette dégradation affecte moins le système SPARSE, qui semble pouvoir mieux modéliser les caractéristiques de la parole mélangées dans le bruit. Pour l'intervalle de RSBs entre 10 dB et 40 dB, les performances de ce dernier sont presque qu'exclusivement supérieures comparativement au système de référence. Pour un RSB de 10 dB, les performances du système SPARSE sont nettement supérieures comparativement à l'entraînement en condition propre, et ce, pour tous les types de bruits. Ceci démontre qu'un entraînement avec bruit est bénéfique à bas-RSB, même s'il affecte légèrement les performances de reconnaissance à haut-RSB. L'usage de caractéristiques parcimonieuses et à grandes dimensions permet de limiter ce problème et d'améliorer la capacité à généraliser.
En effet, une différence majeure est que le système MFCC modélise le bruit à même les dimensions utiles de la parole, car la représentation est dense. 
Le système SPARSE utilise des caractéristiques parcimonieuses et hiérarchiques, et effectue une séparation telle que le bruit sera distribué sur des dimensions différentes de celles de la parole. Dans l'espace des paramètres, le bruit sera alors modélisé avec ses propres fonctions de densité de probabilité, donc dans un sous-espace disjoint où la variabilité sera concentrée hors des dimensions utiles de la parole.

\begin{table}[htb!]
\small
\centering
\caption{Taux de reconnaissance avec entraînement et test en conditions adverses, pour différents types de bruits (babble, destroyerengine, volvo, white)  et RSBs (-5 dB à 40 dB). Les résultats montrent que le système proposé SPARSE permet une meilleure généralisation pour une large gamme de RSBs, comparativement au système de référence MFCC. Il y a aussi une dégradation de base moindre, comparativement à l'entraînement en condition propre. Le bruit durant l'entraînement affecte donc peu les performances absolues.}
\label{tb:perf_train_noisy}

\subtable[Bruit de conversation]{
\begin{tabular}{| r | r | r | r | r | r | r |}
\hline
RSB & -5 dB & 0 dB & 10 dB & 20 dB & 40 dB & Propre\\
\hline \hline
MFCC & \textbf{3.9} & \textbf{10.3} & 34.6 & 55.0 & 59.9 & 41.0  \\
SPARSE & 3.1 & 9.6 & \textbf{63.3} & \textbf{88.8} & \textbf{88.1} & \textbf{87.0} \\
\hline
\end{tabular}
\label{tb:perf_train_noisy_babble}
}

\subtable[Bruit de salle de machine d'un contre-torpilleur (bateau)]{
\begin{tabular}{| r | r | r | r | r | r | r |}
\hline
RSB & -5 dB & 0 dB & 10 dB & 20 dB & 40 dB & Propre\\
\hline \hline
MFCC & \textbf{4.7} & \textbf{8.2} & \textbf{43.5} & 62.9 & 61.5 & 41.0  \\
SPARSE & 2.4 & 3.5 & 37.5 & \textbf{83.1} & \textbf{79.9} & \textbf{87.0} \\
\hline
\end{tabular}
\label{tb:perf_train_noisy_destroyerengine}
}

\subtable[Bruit intérieur d'une voiture]{
\begin{tabular}{| r | r | r | r | r | r | r |}
\hline
RSB & -5 dB & 0 dB & 10 dB & 20 dB & 40 dB & Propre\\
\hline \hline
MFCC & 38.1 & 46.6 & 56.5 & 59.8 & 43.1 & 41.0  \\
SPARSE & \textbf{42.0} & \textbf{67.2} & \textbf{87.6} & \textbf{88.6} & \textbf{87.8} & \textbf{87.0} \\
\hline
\end{tabular}
\label{tb:perf_train_noisy_volvo}
}

\subtable[Bruit blanc gaussien]{
\begin{tabular}{| r | r | r | r | r | r | r |}
\hline
RSB & -5 dB & 0 dB & 10 dB & 20 dB & 40 dB & Propre\\
\hline \hline
MFCC & \textbf{4.6} & \textbf{7.6} & 38.8 & 58.4 & 45.1 & 41.0  \\
SPARSE & 2.2 & 3.4 & \textbf{43.0} & \textbf{84.6} & \textbf{87.0} & \textbf{87.0} \\
\hline
\end{tabular}
\label{tb:perf_train_noisy_white}
}
\end{table}

Un profilage du temps de calcul des différents étages de traitement a été réalisé pour le système de référence et le système proposé, sur un processeur Intel Xeon cadencé à 2.4 GHz (1 coeur). Le tableau \ref{tb:performance_realtime} montre que selon le facteur temps-réel (TR), l'approche proposée SPARSE est beaucoup plus lente que l'approche standard MFCC. Le facteur temps-réel $\tau_{TR} = d_{son}/d_{trait}$ correspond au rapport de la durée $d_{son}$ du son d'entrée sur la durée $d_{trait}$ de traitement (e.g. extraction de caractéristiques, classification). Le cas $\tau_{TR} \geq 1$ indique l'atteinte d'une performance en temps-réel, donc le système peut traiter le son d'entrée en continu (e.g. venant d'un microphone). On cherche à obtenir un facteur temps-réel (TR) le plus haut possible. L'implémentation actuelle de l'architecture proposée ne permet actuellement que le traitement hors-ligne de la parole. 
Plusieurs améliorations au niveau de l'étage d'extraction des caractéristiques permettraient toutefois de réduire le temps de calcul.
Par exemple, lorsque la fréquence d'échantillonnage $F_{sous}$ du vecteur de caractéristiques $\mathbf{Y}_t$ au temps $t$ est choisie telle qu'il y a alignement avec les blocs de projection de premier niveau (i.e. $h=0$), la majorité des projections effectuées au temps $t$ peuvent servir au calcul des bases de niveau $h=0$ et $h=1$ au temps $t+1$. Ceci permet de tirer profit du chevauchement temporel existant entre les vecteurs $\mathbf{Y}_t$ et $\mathbf{Y}_{t+1}$. Enfin, l'évaluation de mixtures de Bernoulli possédant des milliers de paramètres est ce qui alourdit le plus l'étage de classification. 
L'entraînement des systèmes SPARSE et MFCC requiert respectivement 190 minutes et 20 minutes de temps de calcul sur un processeur AMD Opteron cadencé à 2.2 GHz (16 coeurs). L'extraction de caractéristiques et l'évaluation de mixtures de Bernoulli à très grandes dimensions expliquent l'écart considérable entre les temps d'entraînement.

\begin{table}[htb!]
\small
\centering
\caption{Facteur temps-réel (TR) pour l'exécution de l'étage d'extraction des caractéristiques et de classification, ainsi que de l'exécution globale.
Les résultats montrent que la charge de calcul est significativement plus élevée pour l'approche SPARSE, comparativement au système de référence MFCC ayant l'avantage de pouvoir être utilisé en temps-réel.}
\label{tb:performance_realtime}
\begin{tabular}{|r|r|r|r|}
\hline
&Extraction&Classification&Global \\
&($\times$TR)&($\times$TR)&($\times$TR)\\
\hline \hline
MFCC&\textbf{42.437}&\textbf{2.716}&\textbf{2.560}\\
SPARSE&0.792&0.518&0.314\\
\hline 
\end{tabular}
\end{table}

\subsubsection{Effet du bruit sur l'apprentissage des bases}

\begin{figure}[htb!]
\centering
\subfigure[Bases représentant la parole (64 x 160 ms)]{
\begin{tabular}{cccc}
\includegraphics[trim=50 50 50 50, clip, width=0.20\linewidth]{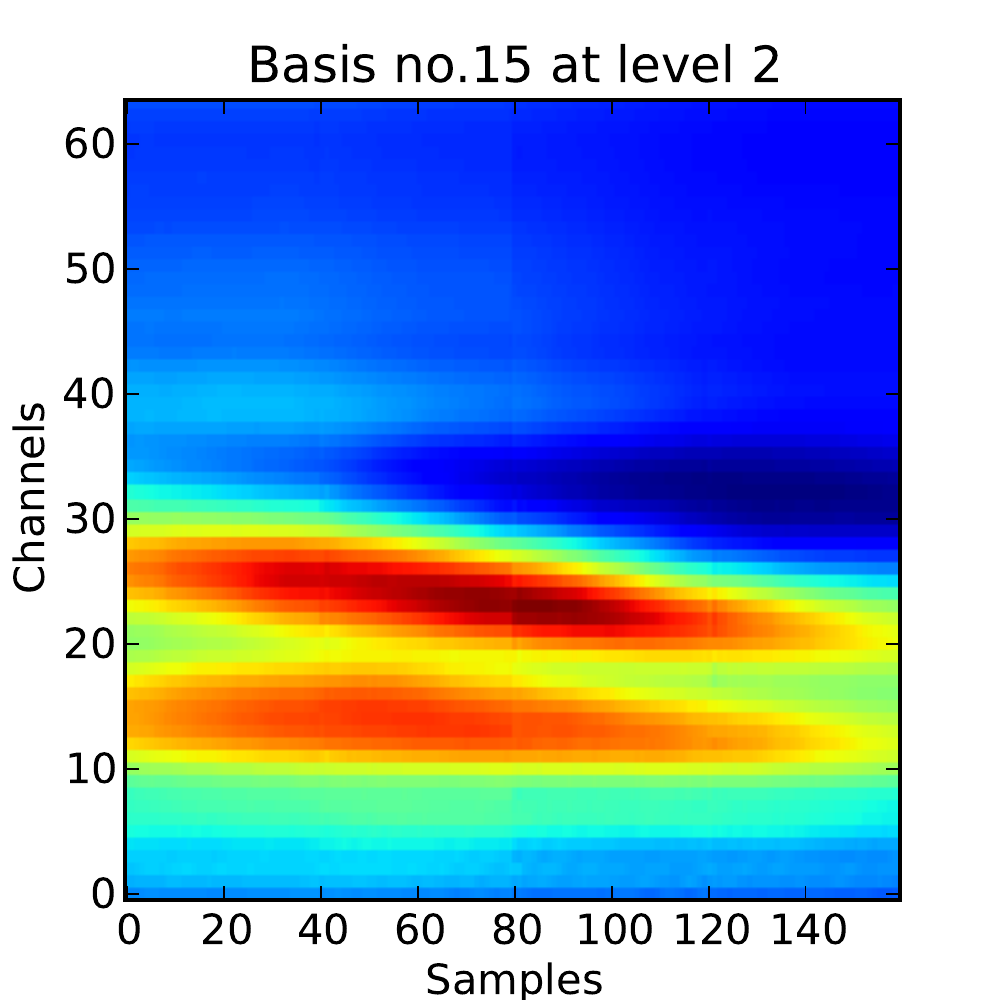}&
\includegraphics[trim=50 50 50 50, clip, width=0.20\linewidth]{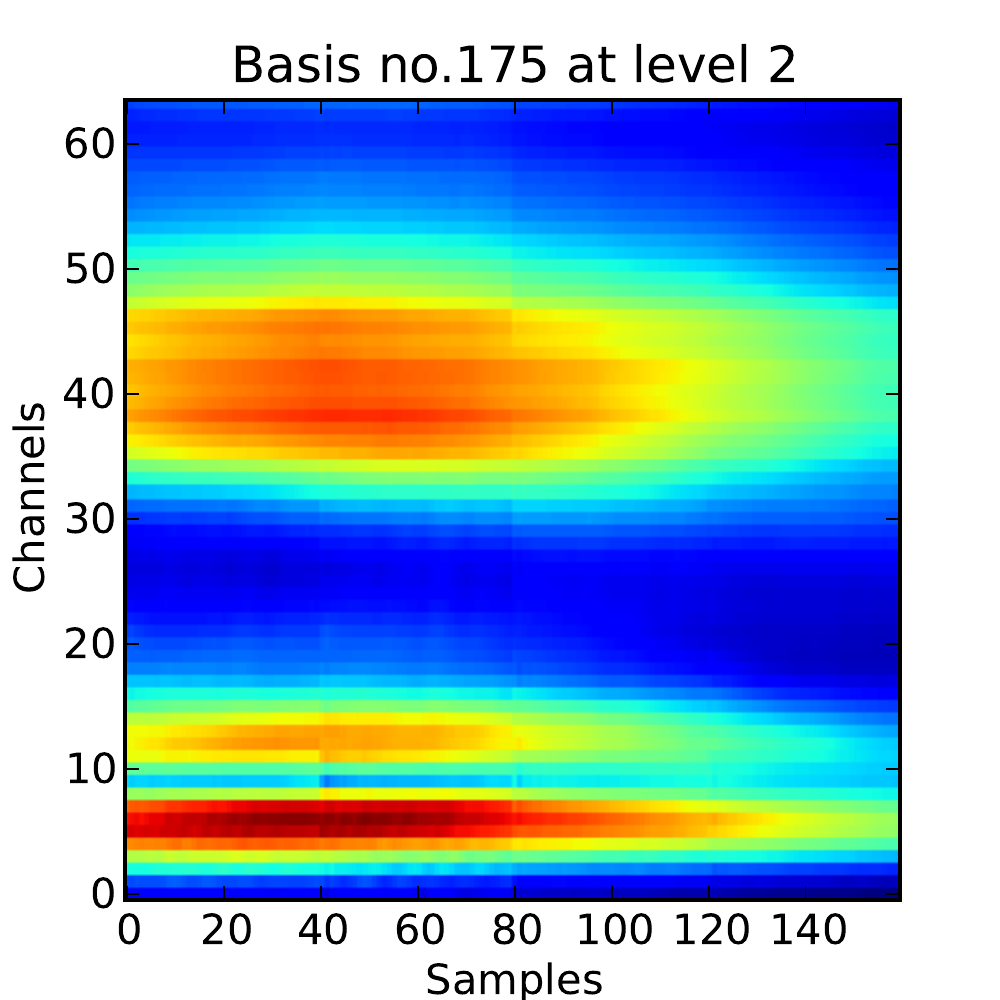}&
\includegraphics[trim=50 50 50 50, clip, width=0.20\linewidth]{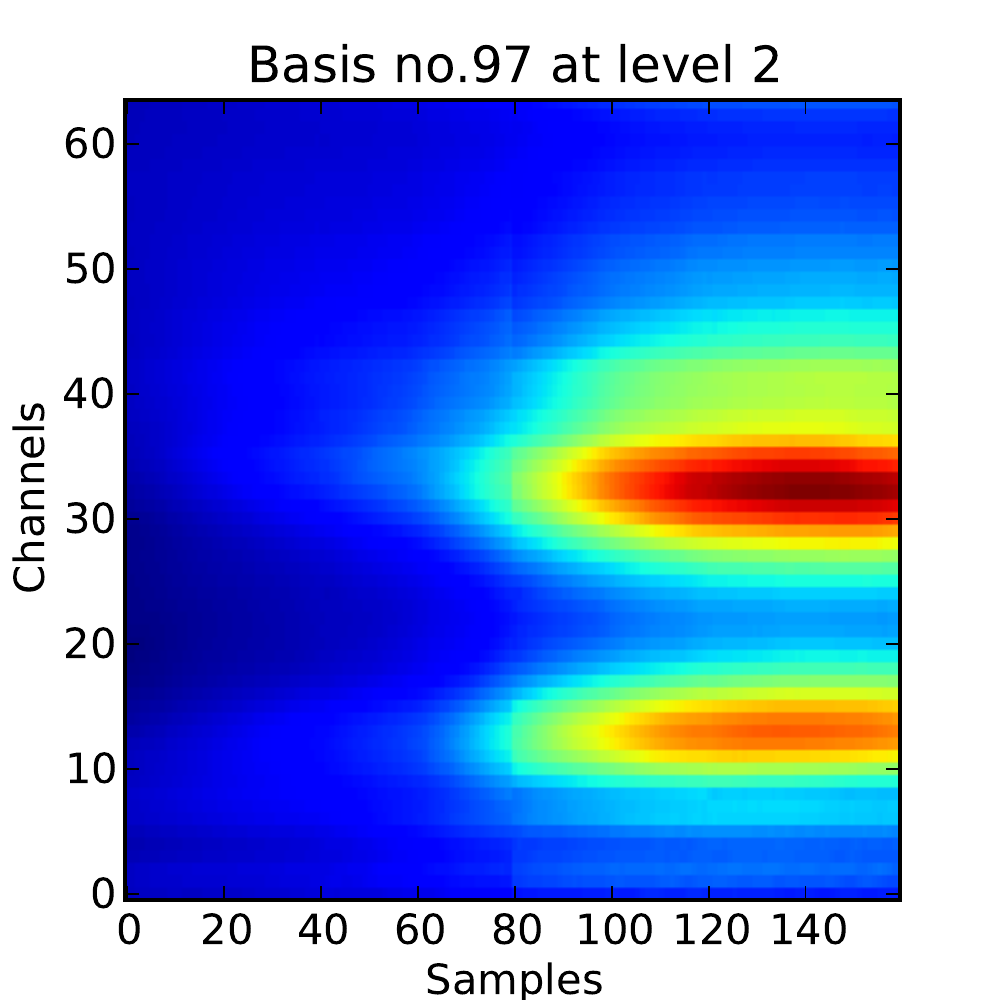}&
\includegraphics[trim=50 50 50 50, clip, width=0.20\linewidth]{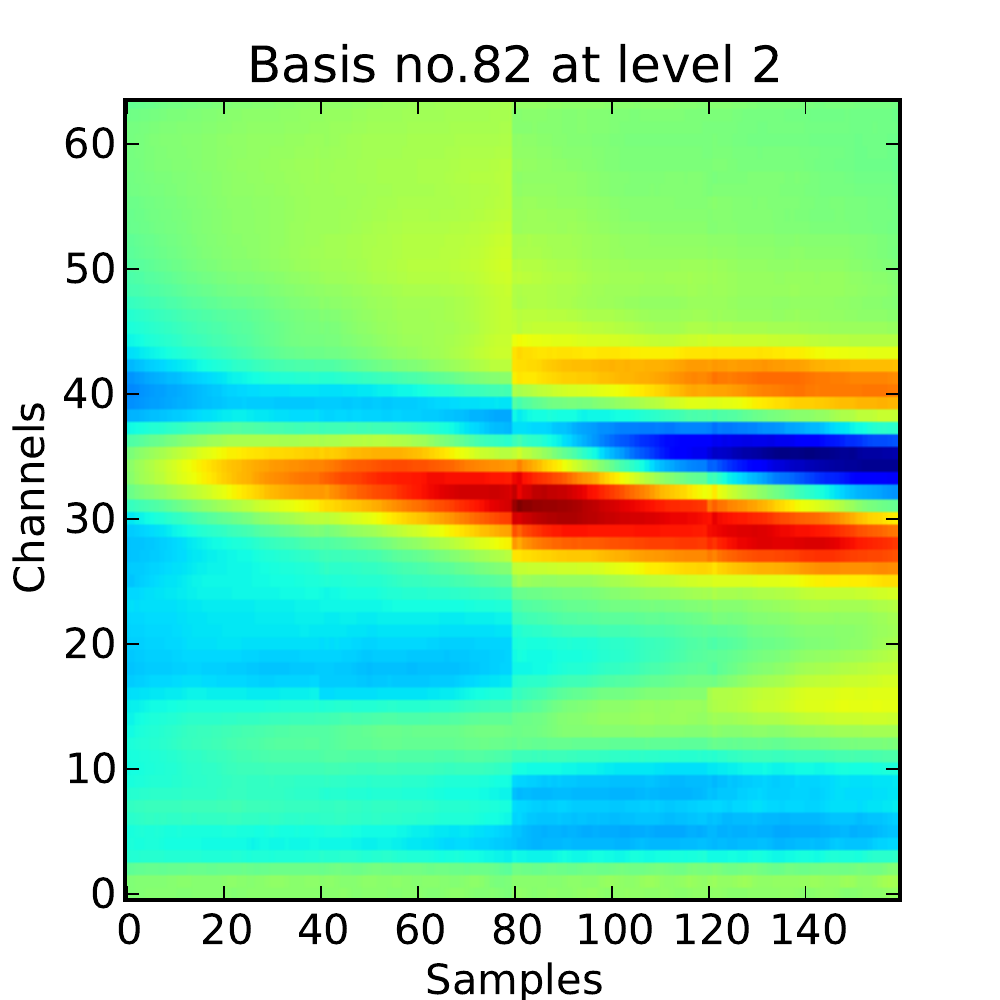}
\end{tabular}
\label{fig:opt_base_speech}
}
\subfigure[Bases représentant le bruit (64 x 160 ms)]{
\begin{tabular}{cccc}
\includegraphics[trim=50 50 50 50, clip, width=0.20\linewidth]{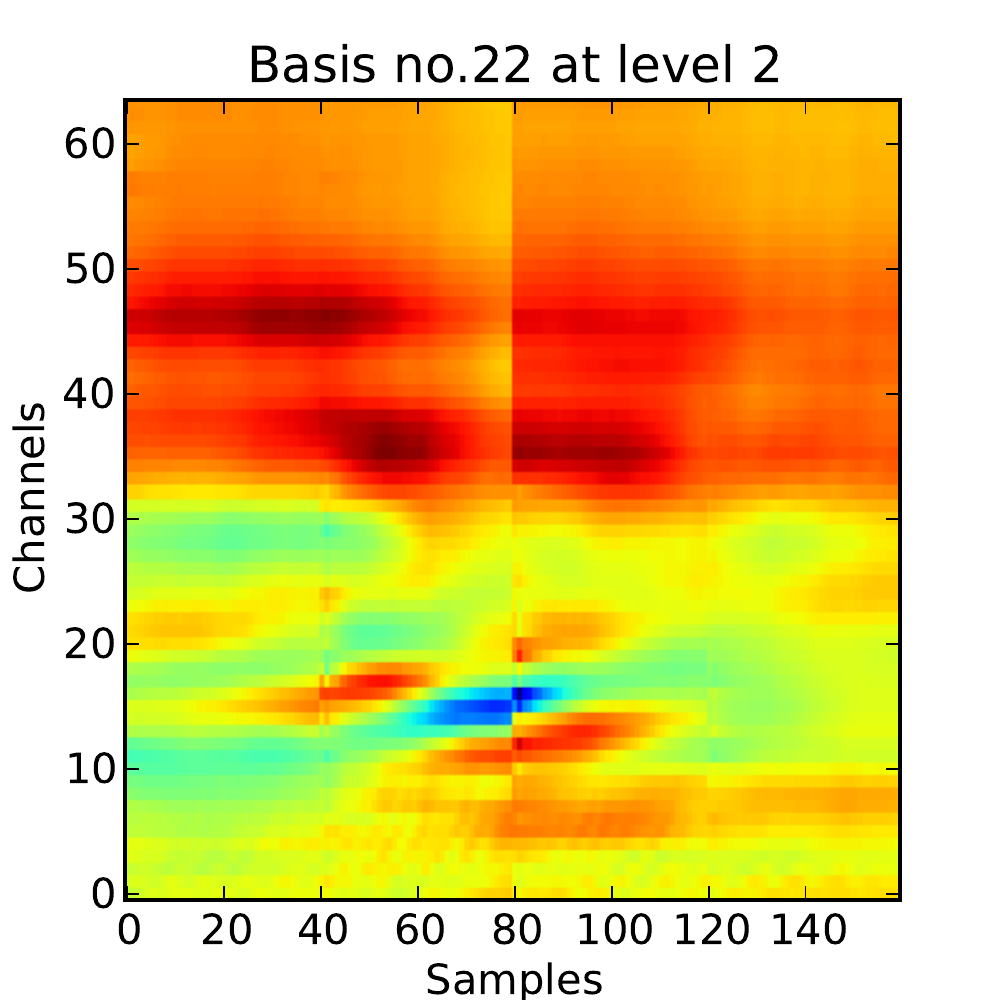}&
\includegraphics[trim=50 50 50 50, clip, width=0.20\linewidth]{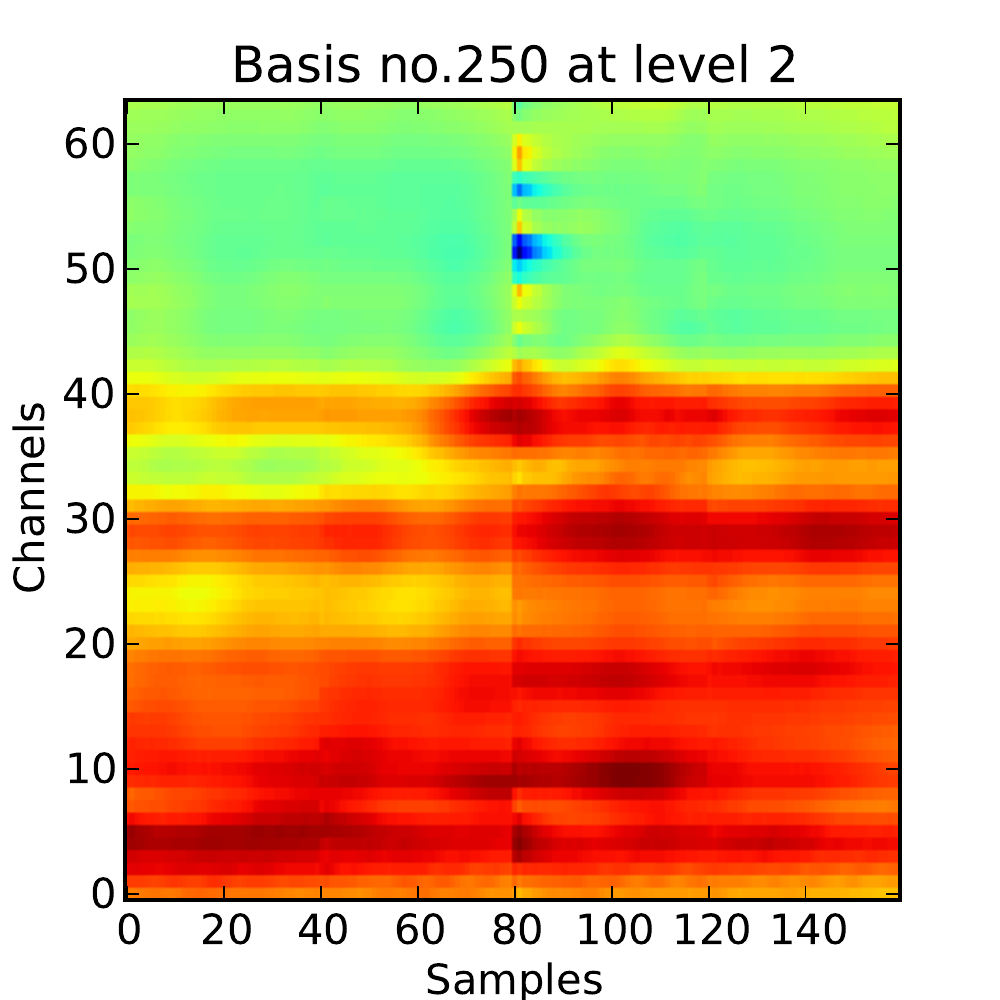}&
\includegraphics[trim=50 50 50 50, clip, width=0.20\linewidth]{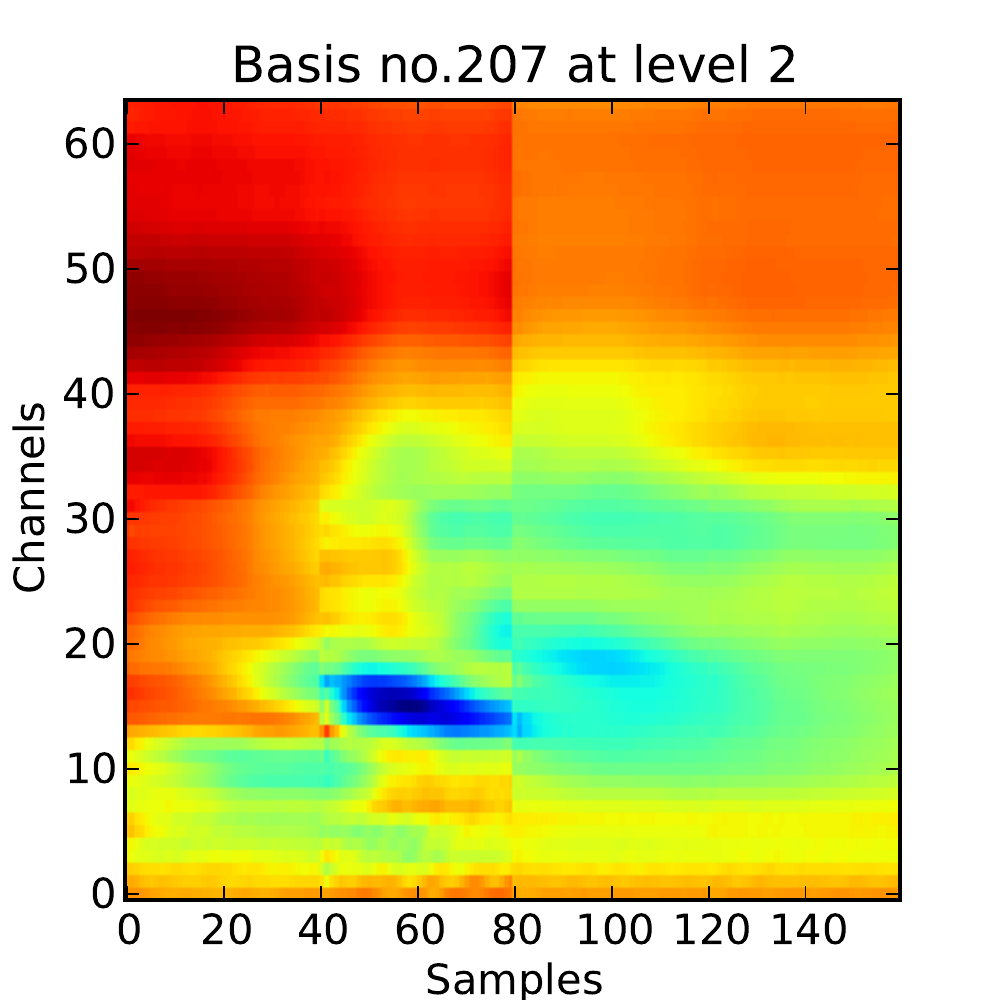}&
\includegraphics[trim=50 50 50 50, clip, width=0.20\linewidth]{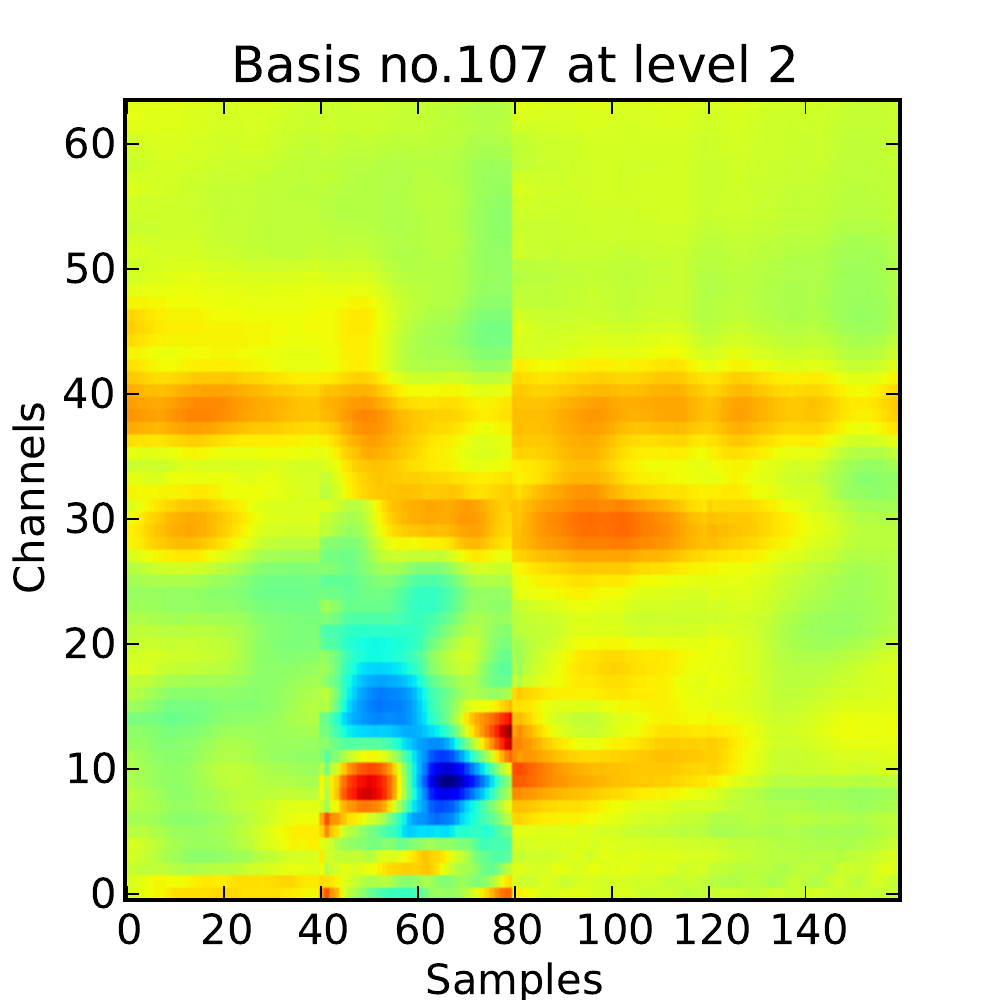}
\end{tabular}
\label{fig:opt_base_noise}
}
\caption{Exemple de bases apprises par analyse en composantes indépendantes (ICA), qui montre une séparation \subref{fig:opt_base_speech} des composantes de la parole et \subref{fig:opt_base_noise} des composantes du bruit. Le non-chevauchement du bruit et de la parole dans l'espace des paramètres du modèle acoustique statistique favorise une dégradation moindre des performances de reconnaissance en conditions adverses.}
\label{fig:opt_base_noise_effet}
\end{figure}

L'analyse en composantes indépendantes (ICA) est couramment utilisée pour la séparation aveugle de sources~\cite[e.g.][]{Obradovic1997}, utile par exemple pour débruiter un signal de parole~\cite[e.g.][]{Lee2000c, Hongyan2010}. Il a déjà été remarqué que l'adaptation des bases en condition de bruit durant l'entraînement est bénéfique~\cite{Heckmann2011}, car on tient alors compte de la variabilité introduite par le bruit. Avec des données d'entraînement bruitées reflétant mieux les conditions de tests, de meilleures performances de reconnaissance doivent être attendues, ce qui corrobore les résultats obtenus. La figure \ref{fig:opt_base_noise_effet} montre qu'effectivement, lorsque les bases sont apprises en présence de bruit, il y a séparation des composantes du bruit de celles de la parole lors de la projection hiérarchique.

Les bases qui ont un contexte temporel plus grand sont plus aptes à extraire les composantes de bruit, car les régularités statistiques à long-terme peuvent être mieux capturées. Il s'agit d'une tâche plus difficile à petite échelle, ce qui explique la pertinence d'effectuer une projection hiérarchique pour l'obtention de bases de haut-niveau. Le choix de l'algorithme d'apprentissage des bases n'est qu'un des nombreux facteurs influençant les performances de reconnaissance, et pas nécessairement le plus important~\citep{ODonnell2012}. Dans le cas présent, l'analyse en composantes indépendantes (ICA) dans un contexte de projection hiérarchique s'est révélée l'algorithme d'apprentissage non-supervisé idéal. L'obtention du vecteur de caractéristiques parcimonieux à grandes dimensions est alors faite en s'assurant que les caractéristiques propres à la parole soient fidèlement extraites, même en présence de bruit dans les données d'entraînement. Ceci appuie bien l'aspect de puissance de discrimination des représentations parcimonieuses dans des espaces à grandes dimensions~\cite{Tosic2011}, avec un compromis entre la séparabilité des classes et la qualité d'approximation. Dans le cas de la reconnaissance vocale, l'aspect de discrimination prime, car la reconstruction du signal est sans importance pour son identification.

\section{Discussion}

L'un des objectifs était de comparer la robustesse en conditions difficiles d'un système de reconnaissance de mots isolés basé sur une représentation parcimonieuse à grande dimension. Plusieurs aspects bio-inspirés sont à la base des ces principes, et ont inspiré l'extraction de caractéristiques de modulation spectro-temporelle par décomposition linéaire et hiérarchique.

L'analyse en composantes indépendantes (ICA) semble être l'algorithme d'apprentissage non-supervisé idéal pour produire, avec l'intégration de ces aspects, des caractéristiques parcimonieuses à grandes dimensions où la parole et le bruit sont restreints dans des sous-espaces disjoints.

L'analyse du tableau~\ref{tb:perf_train_clean} montre que le système SPARSE est supérieur lorsque l'apprentissage et la reconnaissance se font en conditions propres. Ceci valide l'intérêt de l'étude, puisque les résultats de départ sont au moins supérieurs à ceux d'un système conventionnel. Par contre, avec un apprentissage toujours en conditions propres, mais une reconnaissance en conditions difficiles (ou avec bruits), le système SPARSE proposé n'augmente que peu les taux de reconnaissance. Il est meilleur pour le bruit de voiture et de conversation, mais moins bon pour le bruit blanc et l'environnement sonore du bateau. Dans la situation de l'apprentissage avec des données propres, les deux systèmes possèdent des capacités similaires à généraliser en conditions de test difficiles -- pour la majorité des bruits réalistes testés. Le système SPARSE est toutefois plus robuste à la présence de bruit dans les données d'entraînement (voir Tableau~\ref{tb:perf_train_noisy}). On observe une dégradation moindre des taux de reconnaissance à des rapports signal-à-bruit entre 10 dB et 40 dB. En entraînement multi-condition, le système SPARSE permet donc de mieux généraliser, car les bases apprises lors de la projection hiérarchique créent une séparation des composantes de parole de celles du bruit. Cette séparation est conservée au niveau du modèle acoustique, soit dans les paramètres des mixtures de Bernoulli à grandes dimensions.

Les représentations parcimonieuses et par objets de la parole promettent donc d'améliorer les taux de reconnaissance des systèmes utilisés dans des environnements non-contrôlés et riches en bruits additifs, principalement là où les données d'entraînement sont de moindre qualité (i.e. aussi légèrement bruitées).
Les travaux futurs porteront à valider l'approche proposée en reconnaissance de parole continue sur une tâche à large vocabulaire, ainsi qu'à augmenter la rapidité d'exécution.

\section{Conclusion}
La définition d'un objet sonore est différente selon qu'on se place du point de vue de la production ("génération") ou de la perception de l'objet. D'un point de vue de la production du son, on se plait à concevoir qu'un objet sonore est composé de caractéristiques spécifiques qui se succèderaient dans le temps (e.g. ONSET ou transitoire suivi d'un cours silence, puis d'une voyelle). La synthèse de parole repose souvent sur cette conception~\cite{Keller1994}. En conséquence, la génération de certains objets sonores peut être réalisée par l'établissement d'une succession d'évènements acoustiques qui sont considérés comme étant distincts dans le temps (la qualité perceptive n'est d'ailleurs pas toujours excellente). En revanche la perception d'un objet sonore ne peut s'établir de la même façon. Un objet sonore simple n'est pas perçu comme étant la succession d'évènements discrets élémentaires temporels, mais plutôt comme étant un tout qui est le résultat de l'intégration spatio-temporelle des caractéristiques élémentaires par le système auditif. Par exemple le son /b/ est perçu comme étant un tout et non pas comme étant composé d'une barre de voisement en basse fréquence (activation des cordes vocales)  suivie d'un transitoire puis de la voyelle.
En reconnaissance automatique des sons, il semble donc plus logique de vouloir utiliser le point de vue perceptif d'un objet sonore que le point de vue de la génération de l'objet sonore pour pouvoir atteindre des performances qui soient le plus proche possible de celles de l'humain.

L'approche utilisée dans le présent travail considère un objet sonore comme étant la combinaison spatio-temporelle et hiérarchique d'unités élémentaires. Ces unités élémentaires ont été trouvées par apprentissage non-supervisé suivant l'analyse en composante indépendante (ICA) et une organisation hiérarchique. Le critère de recherche de ces unités élémentaires qui a été choisi repose essentiellement sur des données physiologiques qui montrent l'existence d'indépendance statistique entre certains neurones corticaux, permettant ainsi de maximiser la quantité d'information (entropie) par neurones ou groupes de neurones. Nous observons qu'avec le critère choisi, les parties d'objet sonores (ou bases) sont des combinaisons de patrons de transitoires, de modulations en amplitude (AM) et en fréquence (FM) des signaux d'enveloppes extraits à partir d'un banc de filtres cochléaires. Par ailleurs, la recherche automatique et non-supervisée de ces unités élémentaires permet une adaptation de l'approche à différents contextes acoustiques.

Toutefois, ce travail ne répond que très partiellement aux questions posées à l'introduction de cet article. En effet, la façon de combiner les caractéristiques entre elles est arbitraire (utilisation d'un seul vecteur) et seule l'information des enveloppes à la sortie d'un banc de filtres cochléaire a été exploitée. Il est reconnu que le système auditif est aussi en mesure de résoudre les harmoniques en basse fréquence, car un nombre important de fibres basses fréquences du nerf auditif (et donc de cellules ciliées de la cochlée) déchargent selon la période de la fondamentale du signal et non pas selon l'enveloppe des sorties du banc de filtres~\cite{Miller1984}. Cet aspect n'a pas été pris en compte dans le présent travail, car l'objectif n'était pas d'intégrer les caractéristiques de suivi de fréquence fondamentale ou de prosodie dans la représentation des objets sonores utilisés en reconnaissance.

Il a été illustré comment l'intégration de principes inspirés des neurosciences permet de proposer une représentation objet de signaux sonores. Il aurait été possible de combiner différemment les composantes des  objets sonores. En effet, la façon de combiner affecte la représentation des objets sonores et donc les résultats. Ici, il a été choisi d'utiliser une combinaison très simple: chaque dimension d'un vecteur est une caractéristique d'objet (ou de partie d'objet), peu importe sa position ou son rôle dans la structure de l'objet sonore en question. Malgré cette combinaison non ordonnée et non hiérarchisée des composantes des objets, les résultats sont très prometteurs.

Ce travail montre par ailleurs qu'il est possible de tirer profit des représentations à très grandes dimensions. Ces représentations sont rarement considérées comme intéressantes en modélisation statistique des signaux, notamment en raison de la complexité anticipée. Cependant, la complexité de calcul peut être grandement réduite, car l'extraction des caractéristiques reste simple et se prête bien au calcul distribué. De plus, les vecteurs des composantes objets sont à coordonnées binaires. Ceci permet aussi une grande simplicité de mise en oeuvre.

\section*{REMERCIEMENTS}
Calcul Canada, pour les ressources de calcul de haute-performance mises à disposition. Le conseil de recherches en sciences naturelles et en génie du Canada (CRSNG), le fonds de recherche Québec en Nature et Technologies (FRQ-NT).


\end{document}